\documentclass[twocolumn,eqsecnum,showpacs,aps,amsmath,amssymb,prb]{revtex4}
\usepackage{bm}

\newcommand{\leftact}{\overset{\rightarrow}}
\newcommand{\rightact}{\overset{\leftarrow}}
\newcommand{\leftrightact}{\overset{\leftrightarrow}}
\newcommand{\s}{\mbox{\tiny S}}
\newcommand{\B}{\mbox{\tiny B}}
\newcommand{\ti}{\Tilde}

\newcommand{\nl}{\nonumber \\}
\newcommand{\nla}{\nl&\quad}

\newcommand{\Sec}[1]{Sec.\;\ref{#1}}
\newcommand{\App}[1]{Appendix\;\ref{#1}}
\newcommand{\be}{\begin{equation}}
\newcommand{\ee}{\end{equation}}
\newcommand{\bea}{\begin{eqnarray}}
\newcommand{\eea}{\end{eqnarray}}
\newcommand{\bsube}{\begin{subequations}}
\newcommand{\esube}{\end{subequations}}
\newcommand{\Eq}[1]{Eq.\,(\ref{#1})}
\newcommand{\Eqs}[1]{Eqs.\,(\ref{#1})}

\newcommand{\dg}{\dagger}
\newcommand{\la}{\langle}
\newcommand{\ra}{\rangle}

\newcommand{\ind}{{\sf n}}

\begin{document}

\title{Exact dynamics of  dissipative electronic systems
   and quantum transport: \\ Hierarchical equations of motion approach}

\author{Jinshuang Jin}
\author{Xiao Zheng}
\author{YiJing Yan}

\affiliation{Department of Chemistry, Hong Kong University
   of Science and Technology, Kowloon, Hong Kong}

\date{12 October 2007}%, submit to PRB; MS\#BK10673; revised 4 December 2007}

\begin{abstract}
   A quantum dissipation theory is formulated
 in terms of hierarchically coupled equations of motion
 for an arbitrary electronic system coupled with
 grand canonical Fermion bath ensembles.
  The theoretical construction starts with
 the second--quantization influence functional in path integral
 formalism, in which the Fermion creation
 and annihilation operators are represented by
 Grassmann variables. Time--derivatives on
 influence functionals are then performed
 in a hierarchical manner, on the basis of
 calculus--on--path--integral algorithm.
 Both the multiple--frequency--dispersion and
 the non-Markovian reservoir parametrization schemes
 are considered for the desired hierarchy construction.
 The resulting formalism is in principle exact, applicable
 to interacting systems, with arbitrary
 time-dependent external fields.
 It renders an exact tool to evaluate
 various transient and stationary quantum transport properties
 of many-electron systems.
  At the second--tier truncation level the present
 theory recovers
 the real--time diagrammatic formalism
 developed by Sch\"{o}n and coworkers.
  For a single-particle system, the hierarchical formalism
 terminates at the second tier exactly,
 and the Landuer--B\"{u}ttiker's transport
 current expression is readily recovered.
\end{abstract}

\maketitle

\section{Introduction}
\label{thintro}
  The aim of this work is to establish
 an exact quantum dissipation theory (QDT),
 with which various quantum transport properties
% for arbitrary electronic (or spintronic) systems
 could in principle be evaluated without approximations.
 Quantum transport through nanosystems has
 conventionally been studied via
 the Landauer--B${\rm\ddot{u}}$ttiker scattering theory\cite{Dat95}
 and nonequilibrium Green's function (NGF) formalism.\cite{Hau96}
 These approaches are however basically
 single-particle theories.
  The fundamental physics of quantum transport through an
 interacting system has been understood
 based on methods applicable to certain limits;
 for example, in either the weak or strong Coulomb interaction
 regime.\cite{Her913720,Yey932991,Fuj03155310,Hau96,Mei913048,Mei932601}
%%%
 Sch\"{o}n and coworkers have developed a
 formulation based on real--time
 diagrammatic technique.\cite{Sch9418436,Sch94423,%
 Kon9531,Kon9616820,Kon961715}
 This approach could in principle be exact; however,
 to identify all diagrams is itself formidable,
 and there are practically no feasible ways to overcome
 this difficulty for a general non-Markovian system
 in the presence of arbitrary Coulomb interaction.
  Other promising nonperturbative methods,
 such as the numerical renormalization
 group approach,\cite{Wil75773,Kri801044,Hew93}
 are yet to be extended to dynamical quantum systems.

  Quantum transport is studied with the situation,
 where a ``device'', such as a semiconductor quantum dot
 or organic molecule under investigation,
 is connected to electrodes under applied bias voltages.
 This situation can be well described
 in the framework of QDT.  The latter
 concerns the fundamental formulation
 that governs the dynamics of a quantum open system.
  The primary quantity in QDT is the reduced system
 density operator, $\rho(t) \equiv {\rm tr}_{\B}\rho_{\rm T}(t)$.
 Here, $\rho_{\rm T}(t)$ denotes the total density operator
 of system--bath composite; tr$_{\B}$ the
 partial trace over electrode bath degrees of freedom.
 %%%%%%%%%
   Quantum transport based on QDT approach
 has been formulated extensively,\cite{Bru941076,Bru974730,Leh02228305,%
 Li05205304,Wel06044712,Har06235309,Li07075114,Cui06449}
 including the aforementioned real--time diagrammatic
 formalism.\cite{Sch9418436,Sch94423,Kon9531,Kon9616820,Kon961715}
 This approach has the advantage of its generality, since
 different scattering processes can be handled in a
 unified manner,
 and transient dynamics can be studied readily.
 However, the QDTs used in quantum transport by far
 are all perturbative in nature;
 most of them are only of the second order in the system--bath
 coupling.\cite{Leh02228305,Li05205304,Wel06044712,Har06235309,%
 Li07075114,Cui06449}
  A second--order theory is also
 related to the sequential tunneling regime.
  Other limitations include
 the moderately high temperature/voltage
 and quasi-broadband (or quasi-Markovian) approximation.
 Quantum transport with cotunneling processes\cite{Fra01878,Sch05206805}
 has been studied with the fourth-order QDTs, constructed via
 the standard projection operator technique.\cite{Nak58948,Zwa61,Mor65423}
  A self--consistent Born approximation to nonperturbative QDT
 has also been proposed to recover such as the nonequilibrium
 Kondo effect in a model system.\cite{Cui06449}
  Approximations involved in
 the existing quantum transport theories, on the basis of
 either QDT or NGF formalism, are
 subject to ever increasing challenge,
 due to especially the emerging fields of
 quantum measurement and quantum information
 processes.\cite{Gur9715215,Li04085315,Los98120,Fuj06759}

   This work continues our recent effort on the
 development of  QDT
 formalism.\cite{Yan05187,Xu05041103,Xu07031107,Jin07134113}
 The most relevant one is Ref.\ \onlinecite{Jin07134113},
 in which a nonperturbative theory was constructed
 for open quantum system, interacting with
 Boson-like grand canonical bath ensembles.
    The present paper exploits
 the second--quantization field theory
 that properly treats fermionic transfer
 coupling processes.
 % via Grassmann algebra.%\cite{Ryd96,Kle06}

 The remainder of the paper is organized as follows.
 In \Sec{thham}, we specify the electron transfer coupling
 Hamiltonian and the fluctuation--dissipation theorem
 to be used in the later development.
  In \Sec{thpathA}, we revisit the influence functional
 path--integral formalism, with the
 Fermion field representation that involves
 Grassmann variables.
  The derivation is detailed in \App{thapp_path}.
 Presented in \Sec{thpathB} is
 the QDT--based expression of transient transport current.
  In \Sec{theom0}, we first consider a differential form of QDT,
 by exploiting the calculus--on--path--integral
 (COPI) algorithm.\cite{Xu05041103,Xu07031107,Jin07134113,%
 Tan89101,Tan906676,Tan06082001}
 We then develop a multi--frequency--dispersed
 hierarchical equations of motion (MFD-HEOM) formalism of QDT.
 The final MFD-HEOM results are summarized in \Sec{ththeo}.
 The present theory is
 in principle exact,
 applicable to arbitrary quantum
 transport systems.
   It is shown
 in \Sec{thgreen} to recover
 the celebrated Landauer-B\"{u}ttiker's transport current
 expression for single--particle systems.
 The involving reduced single-particle
 density matrix dynamics is derived
 in \App{thapp_green}.
 Moreover, it is demonstrated in \Sec{thtier2} that at the second--tier truncation
 the present MFD-HEOM theory recovers the real--time diagrammatic
 formalism developed by Sch\"{o}n and coworkers for
 Coulomb interaction systems.
 In \Sec{ththeo2}, we present
 an alternative but equivalent HEOM formalism
 via parametrization, with the derivation
 detailed in \App{thapp_theo2}.
 Finally, we conclude this work in \Sec{thsum}.

\section{Stochastic coupling Hamiltonian and quantum
    statistical mechanics}
\label{thham}

\subsection{Stochastic transfer coupling Hamiltonian}
\label{thhamA}

  Consider an electron (or spin) transport setup, in
 which a multi-level system such as
 a semiconductor quantum dot or molecule is in
 contact with electrodes
 (labeled by index $\alpha$). Each electrode
 serves as an electron reservoir and is treated
 as a grand canonical Fermion bath ensemble.
 The total system--electrodes composite Hamiltonian
 assumes
 \bsube \label{HT}
 \be \label{HT0}
   H_{\rm T}= H + \sum_{\alpha}
   \big(h_{\alpha} + H'_{\alpha}\big),
 \ee
 with the noninteracting electrons in the $\alpha$--electrode,
 \be \label{h_alp}
   h_{\alpha}= \sum_{k} \epsilon_{\alpha k }d^\dg_{\alpha k}d_{\alpha k},
 \ee
 and the system--electrode transfer coupling,
 \be \label{Hint}
   H'_{\alpha}= \sum_{\mu,k}t_{\alpha k \mu}
     d^\dg_{\alpha  k}a_{\mu }+ \mbox{H.c.}
 \ee
 \esube
  H.c.\ stands for the Hermitian conjugate;
 ${a^\dg_\mu}$ and $d^\dg_{\alpha  k}$
 ($a_{\mu}$ and $d_{\alpha k}$)
 are the creation (annihilation) operators for
 an electron (or spin) in the specified spin-orbital
 of the system and the $\alpha$-electrode, respectively.
   The system Hamiltonian, $H=H(t;\{a^{\dg}_\mu,a_\mu\})$
 in the right-hand-side (rhs) of \Eq{HT0}, is rather
 arbitrary, including Coulomb interaction
 and coupling with external time--dependent fields
 such as the electric field induced by a pulsed laser.

  Throughout this work, we set $\hbar\equiv 1$ and
  $\beta_{\alpha}\equiv 1/(k_{\B}T_{\alpha})$,
  with $k_{\B}$ the Boltzmann constant
  and $T_{\alpha}$ the
  temperature of  $\alpha$--electrode.
  Electrons in bare electrode in steady state, either
 equilibrium or nonequilibrium,
 are described by the grand canonical density
 operator,
 \be\label{rhoB}
   \rho^{0}_{\alpha}
  =\frac{ e^{-\beta_{\alpha}(h_{\alpha}-\mu_\alpha \hat N_\alpha)} }
  { {\rm tr}_{\B}[e^{-\beta_{\alpha}(h_{\alpha}-\mu_\alpha \hat N_\alpha)}]}.
  \ee
 Here, $\mu_\alpha$ denotes the chemical potential
 of the $\alpha$-electrode in steady state;
 $\hat N_\alpha=\sum_k d^\dg_{\alpha k} d_{\alpha k}$
 is the particle number operator of
 electrons in $\alpha$--electrode. It satisfies
 $[\hat N_\alpha, h_{\B}]=0$, where $h_{\B}\equiv\sum_{\alpha}
 h_{\alpha}$.
 Denote also
 \be \label{bathaver}
   \la\hat O\ra_{\B}\equiv
     {\rm tr}_{\B} \big(\hat O\rho^{0}_{\B}\big);
  \quad\text{with}\quad
     \rho^{0}_{\B}=\prod_{\alpha}\rho^{0}_{\alpha}.
 \ee

  To describe the stochastic nature of
 the transfer coupling,
 consider \Eq{Hint} in the reservoir $h_{\B}$-interaction picture:
 \be \label{Hprimt}
  H'_\alpha(t)= \sum_{\mu}\hat f^{\dg}_{\alpha\mu}(t)a_{\mu }+\mbox{H.c.},
 \ee
 with
 \be \label{falpt}
   \hat f^\dg_{\alpha \mu}(t)\equiv e^{ih_{\rm B}t}\Big[
     \sum_kt_{\alpha k\mu} d^\dg_{\alpha  k}\Big]e^{-ih_{\rm B}t},
 \ee
 being the stochastic interaction bath operators.
 They satisfy the Gaussian statistics with the Wick's theorem
 for thermodynamic average.
 Also note that
 $\la  \hat f^{\dg}_{\alpha\mu}(t)  \ra_{\B}
 =\la \hat f^{\dg}_{\alpha \mu}(t)\hat f^{\dg}_{\alpha \nu}(\tau) \ra_{\B}=0$,
 and $\la\hat f^{\dg}_{\alpha \mu}(t)\hat f_{\alpha' \nu}(\tau) \ra_{\B}=0$
 if $\alpha\neq \alpha'$. As results,
 the effects of reservoirs on the reduced system
 can be completely determined by
 the two-time correlation functions,
\bsube \label{FFCorr}
 \begin{align}
     C^{+}_{\alpha\mu\nu}(t-\tau)
  &=
    \la\hat f^{\dg}_{\alpha\mu}(t)\hat f_{\alpha \nu}(\tau) \ra_{\B},
 \\
   C^{-}_{\alpha \mu\nu}(t-\tau)
 &=
    \la\hat f_{\alpha \mu}(t)\hat f^{\dg}_{\alpha \nu}(\tau) \ra_{\B} .
 \end{align}
\esube
 It follows immediately the time--reversal symmetry and the
 detailed--balance relations:\cite{Jin07134113}
  \be\label{ctsym}
   [C^{\pm}_{\alpha  \mu\nu}(t)]^{\ast}
   = C^{\pm}_{\alpha  \nu\mu}(-t)
   =e^{\pm\beta_{\alpha}\mu_\alpha}
     C^{\mp}_{\alpha \mu\nu}(t-i\beta_{\alpha}).
  \ee

  Physically, $C^{+}_{\alpha  \mu\nu}(t)$
 describes the processes of electron
 tunneling from the reservoir $\alpha$
 into the system, while $C^{-}_{\alpha  \mu\nu}(t)$
 describes the reverse events. Apparently,
 the reservoir correlation functions are diagonal with respect
 to spin indices; i.e., $C^{\pm}_{\alpha\mu\nu}(t)=0$, if
 $\mu$ and $\nu$ belong to different spins.
 By far, the reservoir states are
 assumed to be time--independent.
 The resulting correlation functions defined in \Eq{FFCorr}
 satisfy the stationary condition of
 $C^{\sigma}_{\alpha\mu\nu}(t,\tau)=C^{\sigma}_{\alpha\mu\nu}(t-\tau)$.
  Nonstationary correlations
 for the case of time--dependent chemical potentials
 applied on electrodes will be considered in
 \Sec{thhamC}; see \Eq{corr2}.

 \subsection{Fluctuation--dissipation theorem}
 \label{thhamB}

 For sake of bookkeeping, we shall hereafter
 use also $\sigma$ to label the $+$ or $-$,
 while $\bar\sigma\equiv -\sigma$
 the opposite sign of $\sigma$. Thus, \Eq{ctsym} can be recast as
 \be \label{ctsym_sig}
   [C^{\sigma}_{\alpha  \mu\nu}(t)]^{\ast}
   = C^{\sigma}_{\alpha \nu\mu}(-t)
   =e^{\sigma\beta_{\alpha}\mu_\alpha}
    C^{\bar\sigma}_{\alpha  \mu\nu}(t-i\beta_{\alpha}).
 \ee
 Denote also $a^{\sigma}_{\mu}$
 for either the creation ($a^{+}_{\mu}\equiv a^{\dg}_{\mu}$)
 or the annihilation ($a^{-}_{\mu}\equiv a_{\mu}$)
 operator for the specified spin-orbital state
 of system.

  Introduce now the spectrum functions
 $\Gamma^{\sigma}_{\alpha\mu\nu}(\omega)$ via
 \be \label{Gam_def}
   C^{\sigma}_{\alpha\mu\nu}(t)
  \equiv \int^{\infty}_{-\infty}\!\!d\omega\,
   e^{\sigma i\omega t} \Gamma^{\sigma}_{\alpha\mu\nu}(\omega).
 \ee
 The above definition is consistent with the fact
 that the correlation functions defined in \Eq{FFCorr}
 are of $[C^{+}_{\alpha\mu\nu}(t)]^{\ast}=
    -i\Sigma^{<}_{\alpha\mu\nu}(t)$
 and $C^{-}_{\alpha\mu\nu}(t)=i\Sigma^{>}_{\alpha\mu\nu}(t)$,
 in relation to the self--energy functions in the Green's function
 technique.\cite{Hau96}
 The present index scheme is however
 more convenient in the construction of QDT
 formalism.
 The frequency-domain counterparts of \Eq{ctsym_sig} are
  \be\label{cwsym}
   \Gamma^{\sigma}_{\alpha  \mu\nu}(\omega)
   = \big[\Gamma^{\sigma}_{\alpha  \nu\mu}(\omega)\big]^\ast
   =e^{-\sigma\beta_{\alpha}(\omega-\mu_\alpha)}
     \Gamma^{\bar\sigma}_{\alpha  \nu\mu}(\omega).
 \ee
 The spectrum functions
 are of positivity,\cite{Yan05187,Jin07134113}
 satisfying
 $\Gamma^{\sigma}_{\alpha  \mu\mu}(\omega)\geq 0$
 and
 $\Gamma^{\sigma}_{\alpha  \mu\mu}(\omega)
  \Gamma^{\sigma}_{\alpha  \nu\nu}(\omega)
  \geq |\Gamma^{\sigma}_{\alpha  \mu\nu}(\omega)|^2$.

  To express the detailed--balance relation in terms of
 fluctuation-dissipation theorem (FDT), let us consider
 the interaction spectral density functions, defined as
 \be\label{Jw_def}
   J_{\alpha\mu\nu}(\omega)
  \equiv \frac{1}{2\pi} \int_{-\infty}^{\infty}\!\!dt\,
   e^{i\omega(t-\tau)}
    \la \{\hat f_{\alpha\mu}(t),
          \hat f^{\dg}_{\alpha\nu}(\tau)\}
    \ra_{\B}.
 \ee
 For the present model of linear coupling with noninteracting
 bath reservoir, it can be evaluated as
 $ J_{\alpha\mu\nu}(\omega) = \sum_{k}
   t^\ast_{\alpha k \mu} t_{\alpha k\nu}
   \delta(\omega - \epsilon_{\alpha k})$.
 Together with \Eqs{FFCorr}, (\ref{Gam_def}), and
 the first identity in \Eq{cwsym}, we have
 \be \label{Jwsym}
   J_{\alpha\mu\nu}(\omega)
  =
  \Gamma^{-}_{\alpha\mu\nu}(\omega)+
   \Gamma^{+}_{\alpha\nu\mu}(\omega)
  =
   J^{\ast}_{\alpha\nu\mu}(\omega).
 \ee
 The detailed--balance relation
 [i.e.\ the second identity in \Eq{cwsym}]
 leads then to
 \bsube \label{FDTw}
 \begin{align}
    \Gamma^{+}_{\alpha\mu\nu}(\omega)
 &=
   f^{+}_{\alpha}(\omega) J_{\alpha\nu\mu}(\omega),
 \label{FDTw1} \\
   \Gamma^{-}_{\alpha\mu\nu}(\omega)
 &=
   f^{-}_{\alpha}(\omega)J_{\alpha\mu\nu}(\omega),
 \label{FDTw2}
 \end{align}
 \esube
 with $f^{+}_{\alpha}(\omega)=1-f^{-}_{\alpha}(\omega)
    \equiv f_{\alpha}(\omega)$
 being the Fermi distribution function; i.e.,
 \be\label{fermi_dis}
   f^{\sigma}_{\alpha}(\omega)\equiv
  \frac{1}{1+e^{\sigma\beta_{\alpha}(\omega-\mu_{\alpha})}}.
 \ee
 By setting $J^{+}_{\alpha\mu\nu}\equiv J_{\alpha\nu\mu}$
 and $J^{-}_{\alpha\mu\nu}\equiv J_{\alpha\nu\mu}$,
 one can write \Eq{FDTw} as
 \be\label{FDT}
   C^{\sigma}_{\alpha\mu\nu}(t)
  = \int_{-\infty}^{\infty}\!d\omega
   \frac{ e^{\sigma i\omega t} J^{\sigma}_{\alpha\mu\nu}(\omega)
       }{1+e^{\sigma\beta_{\alpha}(\omega-\mu_{\alpha})}}.
 \ee
 This is the FDT in the fermionic
 grand canonical ensembles.
 It relates the correlation functions
 to the spectral densities.

\subsection{Correlation functions in the presence of time-dependent
 chemical potentials}
\label{thhamC}
  We shall also be interested in the
 transient dynamics of charge transport
 under time--dependent bias voltage.
 Its effect can be
 described by rigid homogeneous
 time--dependent shifts of the conduction
 bands of electrodes; i.e.,
  ${\epsilon}_{\alpha k\mu}(t)
  = \epsilon_{\alpha k\mu} + \Delta_\alpha(t)$
  and
  ${\mu}_\alpha(t) = \mu_\alpha + \Delta_\alpha(t)$,
  so that the occupation on each state
 is unchanged.\cite{Li07075114}
 As results, the nonstationary correlation functions
 are
 \be \label{corr2}
    {C}^{\sigma}_{\alpha\mu\nu}(t,\tau)
  = \exp\left[ \sigma i\! \int_\tau^t\!dt'\Delta_\alpha({t'}) \right]
     {C}^{\sigma}_{\alpha\mu\nu}(t-\tau),
 \ee
 for $t\ge\tau$. Note that $\Delta_{\alpha}(t)$ in this work
 represents time--dependent chemical potential,
 on top of the constant part $(\mu_{\alpha}-\mu^{\rm eq}_{\alpha})$
 applied on $\alpha$--electrode.

\section{Quantum transport versus dissipation}
\label{thpath}

 \subsection{Influence functional in path integral formalism}
 \label{thpathA}

   We shall be interested in a QDT-based transport
 formulation. Let us first revisit the path--integral (PI)
 influence functional expression.
 It serves as the starting point for the
 development of HEOM formalism of QDT.\cite{Tan89101,%
  Tan906676,Tan06082001,Xu05041103,Jin07134113,Xu07031107}
 The quantity of interest here is the
 reduced density operator, defined as the
 trace of the total density operator
 over the bath subspace; i.e.,
  $\rho(t) \equiv {\rm tr}_{\B} \rho_{\rm T}(t)$.
  The quantum dissipation starts with the initial
 condition that the system and bath were initially
 uncorrelated:
 \be\label{rhoT0}
    \rho_{\rm T}(t_0) = \rho(t_0)\rho^0_{\B};
 \qquad t_0\rightarrow -\infty.
 \ee
  This  initial factorization ansatz does not cause
 approximation, as long as the initial time is set
 to infinite past.\cite{Wei99,Xu029196,Yan05187}
 We will come back to this issue in \Sec{thsum}.

   The key variation from our previous work
   with Boson-like reservoir\cite{Jin07134113}
 is the second quantization that leads to the
 Grassmann variables for Fermion fields rather than
 the ordinary c-numbers in the PI representation.\cite{Ryd96,Kle06}
%% %The details are as follows.
%%%%%
  To proceed, let $\{|\psi\ra\}$ be a second-quantization basis
 set in the system subspace, and
 $\bm\psi \equiv (\psi,\psi')$ for short,
 so that $\rho(\bm\psi,t)\equiv \rho(\psi,\psi',t)$.
  Denote ${\cal U}(t,t_0)$ as the reduced Liouville--space propagator,
 by which
 \bsube \label{U0_def}
 \be\label{U0_defA}
   \rho(t) \equiv {\cal U}(t,t_0)\rho(t_0).
 \ee
  Its PI expression in the $\psi$-representation reads\cite{Fey63118}
 \be \label{calU0}
   {\cal U}(\bm\psi,t;\bm\psi_0,t_0)
 = \int_{\bm\psi_0[t_0]}^{\bm\psi[t]}   \!\!  {\cal D}{\bm\psi} \,
     e^{iS[\psi]} {\cal F}[\bm\psi ] e^{-iS[\psi']}.
 \ee
 \esube
 $S[\psi]$ is the classical action functional
 of the reduced system, evaluated along a path $\psi(\tau)$, with the
 constraints that two ending points $\psi(t_0)=\psi_0$ and
 $\psi(t)=\psi$ are fixed.

  The key quantity in PI expression is the
 influence functional, ${\cal F}$ in \Eq{calU0}.
 For the present model of linear coupling with
 noninteracting electron reservoir, it can be
 formally evaluated by using the Wick's theorem
 for the thermodynamic Gaussian average,
 implemented with Grassmann algebra.\cite{Ryd96,Kle06}
 The final results read [cf. \Eq{app_calR}]
 \bsube\label{calF_R}
 \be\label{calF0}
  {\cal F}[\bm\psi] = \exp\left\{-\!\int^t_{t_0}d\tau{\cal R}
  \big[\tau;\{\bm\psi\}\big]\right\},
  \ee
 with (denoting
  $a^+_{\mu}\equiv a^{\dg}_{\mu}$ and  $a^-_{\mu}\equiv a_{\mu}$)
%%%%%%%%
 \begin{align}\label{calR}
  {\cal R}\big[t;\{\bm\psi\}\big]
 \equiv
    i\sum_{\alpha,\mu,\sigma} {\cal A}^{\bar\sigma}_{\mu}[\bm\psi(t)]
   {\cal B}^{\sigma}_{\alpha\mu}\big(t;\{\bm\psi\}\big).
 \end{align}
 \esube
%%%%%%%%%%%%%%%%%%%%%%
 Here, ${\cal A}^{\sigma}_{\mu}$ and ${\cal B}^{\sigma}_{\alpha\mu}$ are
 Grassmann variables, defined as
  \be \label{calAs}
  {\cal A}^{\sigma}_{\mu}[\bm\psi(t)] \equiv
  a^{\sigma}_\mu[\psi(t)]+ a^{\sigma}_\mu[\psi'(t)],
 \ee
 and
 \be \label{calBs}
  {\cal B}^{\sigma}_{\alpha\mu}(t;\{\bm\psi\})
  \equiv -i[ B^{\sigma}_{\alpha\mu}(t;\{\psi\}) -
   {B}_{\alpha\mu}^{\prime\sigma}(t;\{\psi'\})],
 \ee
 with
 \bsube \label{calBs0}
 \begin{align}
   B^{\sigma}_{\alpha\mu}(t;\{\psi\})
 &\equiv
   \sum_{\nu} \int_{t_0}^{t}\!d\tau\,
    C^{\sigma}_{\alpha\mu\nu}(t,\tau) a^{\sigma}_{\nu}[\psi(\tau)],
  \label{calBs0a} \\
   {B}_{\alpha\mu}^{\prime\sigma}(t;\{\psi'\})
 &\equiv
   \sum_{\nu} \int_{t_0}^{t}\!d\tau\,
    C^{\bar\sigma\,\ast}_{\alpha\mu\nu}(t,\tau)
    a^{\sigma}_{\nu}[\psi'(\tau)].
 \label{calBs0b}
 \end{align}
 \esube
%%%
 Note that ${B}_{\alpha\mu}^{\prime\sigma}=
  ({B}_{\alpha\mu}^{\bar\sigma})^{\dg}$ leads to
 ${\cal B}^{\bar\sigma}_{\alpha\mu}=({\cal B}^{\sigma}_{\alpha\mu})^{\dg}$.
 This property corresponds to the Hermitian
 conjugate relation of auxiliary density
 operators that will be discussed later;
 see \Eqs{rho1sym}, (\ref{rhonsym}) or (\ref{rhonsym_para}).
 The interaction reservoir correlation functions
 in \Eq{calBs0} can be nonstationary to support
 the study of transient current under
 time-dependent bias voltage applied to electrodes.

  All variables involved in \Eqs{calAs} and (\ref{calBs})
 stem from second-quantization of fermion operators.
 They are Grassmann variables
 in PI formalism,\cite{Ryd96}
 following the anticommutation relation, such as
 ${\cal B}_{\mu}^{\sigma}a^{\bar\sigma}_{\alpha\mu}
  = -a^{\bar\sigma}_{\mu}{\cal B}_{\alpha\mu}^{\sigma}$.
  Apparently, the dissipation functional
 ${\cal R}$ [\Eq{calR}], as it consists of bi-fermionic
 variables, and the influence functional ${\cal F}$ [\Eq{calF0}]
 remain as ordinary c-numbers.

  The time derivative on ${\cal F}$ reads
  \be \label{dotF0}
  \partial_t {\cal F}=-{\cal R}{\cal F}
 = -i\!\sum_{\alpha,\mu,\sigma}\!{\cal A}^{\bar\sigma}_{\mu}
  {\cal B}^{\sigma}_{\alpha\mu}{\cal F}
   \equiv
  -i\!\sum_{\alpha,\mu,\sigma}\!\!{\cal A}^{\bar\sigma}_{\mu}
    {\cal F}^{\sigma}_{\alpha\mu} .
 \ee
 Introduce here are a set of first-tier {\it auxiliary influence
 functionals} (AIFs),
 \be \label{calF1}
  {\cal F}^{\sigma}_{\alpha\mu} \equiv {\cal B}^{\sigma}_{\alpha\mu}{\cal F}.
 \ee
 They are Grassmann variables, for which [cf.\ \Eq{calAs}]
 \be\label{calAF1}
   {\cal A}^{\bar\sigma}_{\mu}{\cal F}^{\sigma}_{\alpha\mu}
  = a^{\bar\sigma}_\mu[\psi(t)]{\cal F}^{\sigma}_{\alpha\mu}
   - {\cal F}^{\sigma}_{\alpha\mu}a^{\bar\sigma}_\mu[\psi'(t)].
 \ee
 The identity of
 $a^{\bar\sigma}_\mu[\psi'(t)]{\cal F}^{\sigma}_{\alpha\mu}
 = -{\cal F}^{\sigma}_{\alpha\mu}a^{\bar\sigma}_\mu[\psi'(t)]$
 for Grassmann variables is used here.

  The first-tier auxiliary density operators (ADOs) can now be defined via
 [cf.\ \Eq{U0_def}]
\bsube \label{U1_def}
 \be\label{U1_defA}
  \rho^{\sigma}_{\alpha\mu}(t)
   \equiv {\cal U}^{\sigma}_{\alpha\mu}(t,t_0)\rho(t_0),
 \ee
 with
 \be \label{U1_defB}
   {\cal U}^{\sigma}_{\alpha\mu}(\bm\psi,t;\bm\psi_0,t_0)
 \equiv \int_{\bm\psi_0[t_0]}^{\bm\psi[t]}   \!\!  {\cal D}{\bm\psi} \,
     e^{iS[\psi]} {\cal F}^{\sigma}_{\alpha\mu}[\bm\psi ] e^{-iS[\psi']}.
 \ee
\esube
 We can then recast \Eq{dotF0} in terms of the reduced
 density operator and its auxiliary ones as
 \be\label{dotrho0}
 \dot{\rho}(t)=-i{\cal L}\rho(t)-i\sum_{\alpha,\mu,\sigma}
 [a^{\bar\sigma}_{\mu},\rho^{\sigma}_{\alpha\mu}(t)].
 \ee
 The first term
 in which ${\cal L}\rho\equiv[H,\rho]$
 arises from the time derivative of classical action exponential term
 of \Eq{U0_def}. The commutator in the second term can be recast as
 $[a^{\bar\sigma},\rho^{\sigma}_{\alpha\mu}]
  = {\cal A}^{\bar\sigma}_{\mu}\rho^{\sigma}_{\alpha\mu}$ that
 is the operator-level counterpart of \Eq{calAF1}.
 The Hermitian conjugate relation implied in \Eq{calBs} reads
 \be \label{rho1sym}
   \rho^{\bar\sigma}_{\alpha\mu} =
   \big(\rho^{\sigma}_{\alpha\mu}\big)^{\dg}.
 \ee
 This identity can also be verified
 via the equivalent operator-level
 definition of the first-tier ADO
  $\rho^{\sigma}_{\alpha\mu}$; see \Eq{rho1tier}.

 \subsection{Quantum transport current via the auxiliary reduced density
   operator dynamics}
 \label{thpathB}

   In this subsection, we shall show that
 the transport current is directly related to the first-tier
 ADOs.
 The final expression for the transient current
 from $\alpha$-lead to system reads (setting $e=1$)
 \be\label{currI}
   I_{\alpha}(t)
  = i \sum_{\mu} {\rm tr}_{\s} \big\{\rho^{+}_{{\alpha}\mu}(t)
  a_\mu- a^\dg_\mu\rho^{-}_{{\alpha}\mu}(t)\big\}.
 \ee
 Here, $\rho^{+}_{{\alpha}\mu}=\big(\rho^{-}_{{\alpha}\mu}\big)^{\dg}$
 are the aforementioned first-tier ADOs.
 The total current passing through the system
 from L to R electrode
 is then $I(t) = I_{\rm L}(t)-I_{\rm R}(t)$.

  We are now in the position to prove the  current expression,
 \Eq{currI}.
 Let us start with the definition,
 \be \label{curr_def}
   I_{\alpha}(t)=-\frac{d}{dt}\la {\hat N}_{\alpha}\ra
  =i\big\la[\hat N_{\alpha},H_{\rm T}(t)]\big\ra_{\rm T}.
  \ee
 Here, $\la\;\cdot\;\ra_{\rm T} ={\rm Tr}_{\rm T} [\;\cdot\;\rho_{\rm T}(t)]$,
 with
 ${\rm Tr}_{\rm T}\equiv {\rm tr}_{\s}{\rm tr}_{\B}$;
 $\hat N_\alpha$ is the electron number operator
 in the electrode $\alpha$, satisfying
 $[\hat N_{\alpha}, H(t)]=[\hat N_{\alpha},h_{\B}]=0$.
 In \Eq{curr_def},
  $H_{\rm T} = H + H'(t)$ is the
 total composite Hamiltonian given in the
 $h_{\B}$-interaction picture.
 One can readily obtain
  \be\label{curr1}
    I_{\alpha}(t)
 = i\sum_\mu\big\la\big[\hat f^\dg_{\alpha\mu}(t)a_\mu
    -a^\dg_\mu \hat f_{\alpha\mu}(t) \big]\big\ra_{\rm T}.
  \ee
 The equivalence between
 \Eq{curr_def} and \Eq{curr1} can be
 established by recognizing the fact
 that the first-tier ADOs defined in
 the previous subsection can be recast as
  \be\label{rho1tier}
  \rho^\sigma_{\alpha\mu}(t)
  = {\rm tr}_{\B}
  \big[\hat f^\sigma_{\alpha\mu}(t)\rho_{\rm T}(t)\big].
  \ee
 In fact $\dot\rho_{\rm T}(t)=-i[H_{\rm T}(t),\rho_{\rm T}(t)]$
 in this work reads [cf.\ \Eq{Hprimt}]
\[
%  \be\label{dotrhoT}
  \dot\rho_{\rm T}(t)
 = -i{\cal L}\rho_{\rm T}(t)
  -i\sum_{\alpha\mu}\big[\hat f^\dg_{\alpha\mu}(t)a_\mu
  \!+\!
  a^\dg_\mu \hat f_{\alpha\mu}(t),\rho_{\rm T}(t)\big].
%\ee
\]
  Applying now the definition of
  $\rho(t)\equiv{\rm tr}_{\B}\rho_{\rm T}(t)$,
 the trace cyclic invariance, and \Eq{dotrho0},
 it concludes immediately that
 \Eq{rho1tier} does amount to
  the first-tier ADO defined earlier.
 This equivalence can also be proved by the explicit
 PI expression of \Eq{rho1tier}.

   We have therefore established
 the transport current expression [\Eq{currI}],
 in terms of the reduced density operator,
  more precisely the first-tier ADO dynamics.
% %%%%%%%%%%%%%%%%%%%
  Note that the charge current in \Eq{currI} is of the form
 of $I_{\alpha}(t) = \sum_{\mu} I_{\alpha\mu}(t)$.
 It consists of the contributions
 from individual spin-orbitals that couple directly with the lead.
 Physically, it is useful to study individual contributions,
 for example spin--dependent current.

\section{Hierarchical equations of motion approach
   to quantum dissipation}
\label{theom0}

  To complete the formulation, we shall
 develop hierarchically coupled EOM
 for evaluating both $\rho$ and
 its ADOs, in the presence of arbitrary non-Markovian dissipation.
 Both MFD and parametrization schemes of hierarchy will
 be considered.
  The desired theory will be constructed
 via the calculus-on-path-integral (COPI)
 algorithm.\cite{Xu05041103,Xu07031107,Jin07134113,%
 Tan89101,Tan906676,Tan06082001}
 For clarity, we present in this section
 the derivation of MFD-HEOM formalism,
 while summarize its
 final results in the next section.
 The parametrization based HEOM formalism
 will be postponed to \Sec{ththeo2},
 after its MFD equivalence is scrutinized
 in relation with the Landauer--B\"{u}ttiker's
 and the real--time diagrammatic formulations
 of quantum transport.

\subsection{Quantum dissipation theory via calculus on path integrals}
\label{theom0A}

   In the following development we omit the explicit PI
 variables dependence whenever it does not cause confusion.
 Adopt further the abbreviation
 $j =\{\mu\sigma\}$ or $\{\alpha\mu\sigma\}$ whenever applicable.
 So that ${\cal A}_{\bar j}\equiv {\cal A}_{\mu}^{\bar\sigma}$,
 ${\cal B}_{ j}\equiv {\cal B}_{\alpha\mu}^{\sigma}$,
 and \Eq{dotF0} is written as
 \be\label{dotF}
   \partial_t{\cal F} = -i \sum_{j}
    {\cal A}_{\bar j}{\cal B}_{j}{\cal F}.
 \ee
  Note that ${\cal A}_{j}$ [\Eq{calAs}] depends on the
 field at the {\it fixed} ending point $\psi(t)=\psi$ of the path.
 The difficulty of the PI formalism
 is the evaluation of ${\cal B}_{j}$ [\Eq{calBs}] that
 involves $\psi(\tau)$ at all $t_0\leq \tau\leq t$.
 To resolve the memory contained in  ${\cal B}_{j}$,
 consider its time derivative,
 \be\label{dotcalB}
  \partial_t  {\cal B}_{j}=
    \ti{\cal B}_{j} -i \ti {\cal C}_{j} .
 \ee
 Here,
 \bsube \label{ticalBs}
 \be\label{ticalBj}
    {\ti{\cal B}}_{j}\equiv {\ti{\cal B}}^{\sigma}_{\alpha\mu}
 = -i[\ti{B}^{\sigma}_{\alpha\mu}-\ti{B}^{\prime\,\sigma}_{\alpha\mu}],
 \ee
  that is similar to ${\cal B}_{j}$, but with
 [cf.\ \Eqs{calBs} and (\ref{calBs0})]
 \be \label{ticalBs0}
     {\ti B}^{\sigma}_{\alpha\mu}(t;\{\psi\})
  = \sum_{\nu} \int_{t_0}^{t}\!d\tau\,
    {\dot C}^{\sigma}_{\alpha\mu\nu}(t,\tau) a^{\sigma}_{\nu}[\psi(\tau)],
 \ee
 \esube
 and [denoting $C_{\alpha\mu\nu}^{0,\sigma}
   \equiv C_{\alpha\mu\nu}^{\sigma}(0) = C_{\alpha\nu\mu}^{\sigma \ast}(0)$;
  cf.\ \Eq{ctsym}]
 \be \label{calC_path0}
   \ti {\cal C}_{j}
 \equiv \sum_{\nu} \left\{C_{\alpha\mu\nu}^{0,\sigma}
   a^{\sigma}_{\nu}[\psi(t)]
     - C_{\alpha\nu\mu}^{0,\bar\sigma}
     a^{\sigma}_{\nu}[\psi'(t)]\right\} .
 \ee
 Note that $C_{\alpha\mu\nu}^{\sigma}(t,t)
 =C_{\alpha\mu\nu}^{\sigma}(0)$
 used in \Eq{calC_path0} is valid for either stationary or
 nonstationary reservoir correlation functions.
 Apparently, both ${\ti{\cal B}}_{j}$
 and $\ti{\cal C}_{j}$ are Grassmann variables.
 The former contains memory, while the latter
 depends only on the fixed ending point of the path.

  To continue the hierarchy construction, consider now
 the time-derivative on the first-tier AIF of \Eq{calF1}.
 Using \Eqs{dotF} and (\ref{dotcalB}), we have
  (noting that ${\cal F}_{j}\equiv {\cal F}^{\sigma}_{\alpha\mu}$)
 \begin{align}\label{dotF1}
   \partial_t{\cal F}_{j}
    &\equiv
     \partial_t({\cal B}_{j}{\cal F})
 \nl&=
    (\ti{\cal B}_{j}-i\ti{\cal C}_{j}){\cal F}
   -i \sum_{j'}{\cal B}_{j}{\cal A}_{\bar j'}{\cal B}_{j'}{\cal F}
 \nl&=
   \ti{\cal B}_{j}{\cal F}-i\ti{\cal C}_{j}{\cal F}
   -i \sum_{j'}{\cal A}_{\bar j'}{\cal B}_{j'}{\cal B}_{j}{\cal F}
 \nl&\equiv
   \ti{\cal F}_{j}-i\ti{\cal C}_{j}{\cal F}
   -i \sum_{j'}{\cal A}_{\bar j'}{\cal F}_{jj'}.
 \end{align}
 Note that $\ti{\cal F}_{j}\equiv {\ti{\cal B}}_{j}{\cal F}$,
 which is a Grassmann variable, belongs to
 a different class of first-tier AIF. We shall return to this issue
 in the next subsection.

  The last term in \Eq{dotF1} defines also the second-tier AIFs; i.e.,
 \be\label{calF2}
   {\cal F}_{jj'}\equiv {\cal B}_{j'}{\cal B}_{j}{\cal F} = -{\cal F}_{j'j}.
 \ee
 The second identity that implies ${\cal F}_{jj}=0$ arises from
 the Grassmann anticommutation relation of ${\cal B}_{j'}{\cal B}_{j}
 = -{\cal B}_{j}{\cal B}_{j'}$.

  The second-tier AIFs are ordinary c-numbers.
 Thus, ${\cal A}_{\bar j'}{\cal F}_{jj'}$ in the last
 term of \Eq{dotF1} amounts to [noting that
 ${\cal A}_{\bar j'}\equiv {\cal A}^{\bar\sigma'}_{\mu'}$
   of \Eq{calAs}]
 \be \label{calAF2}
   {\cal A}_{\bar j'}{\cal F}_{jj'}
  = a^{\bar\sigma\prime}_{\mu'}[\psi(t)]{\cal F}_{jj'}
    + {\cal F}_{jj'}a^{\bar\sigma\prime}_{\mu'}[\psi'(t)].
 \ee
  It is in contrast to \Eq{calAF1}, which involves
 a first-tier AIF that is a Grassmann variable.

 We are now in the position to summarize
 the above hierarchy construction.
 An $n^{\text{th}}$-tier AIFs can
 be defined in general as
  \be\label{AIF}
    {\cal F}_{\bf j}^{(n)} \equiv
   {\cal F}_{j_1j_2\cdots j_n}\equiv {\cal B}_{j_n}\cdots{\cal B}_{j_2}
     {\cal B}_{j_1}{\cal F}.
 \ee
  Note that exchanging any two indexes in a given AIF
 companies with a ``$-$'' sign. This leads to the property that
 the $n$ indexes in ${\cal F}_{j_1j_2\cdots j_n}$
 should all be different; otherwise, the AIF would be zero.
 Denote also
 \be\label{tiAIF}
 \ti{\cal F}_{\bf j}^{(n)}
 = \big(\ti{\cal B}_{j_n}{\cal B}_{j_{n-1}}\!\!\cdots\!{\cal B}_{j_1}
   + \cdots
   +{\cal B}_{j_n}\!\!\cdots\!{\cal B}_{j_2}\ti{\cal B}_{j_1}\big){\cal F}.
 \ee
 Equations (\ref{dotF}) and (\ref{dotcalB}) lead then to
 \begin{align} \label{dotFn0}
   \partial_t {\cal F}_{\bf j}^{(n)}
  &=
    \ti{\cal F}_{\bf j}^{(n)}
    -i \big( \ti{\cal C}_{j_n}{\cal B}_{j_{n-1}}
      \!\!\cdots\!{\cal B}_{j_1} + \cdots
 \nl&\quad
   +{\cal B}_{j_n}\!\!\cdots\!{\cal B}_{j_2}\ti{\cal C}_{j_1}\big) {\cal F}
   -i {\sum_j}'
    {\cal B}_{j_n}\!\!\cdots\!{\cal B}_{j_1}
     {\cal A}_{\bar j}{\cal B}_{j} {\cal F}
 \nl&=
  \ti{\cal F}_{\bf j}^{(n)}
   -i\sum_{k=1}^{n} (-)^{n-k} \ti{\cal C}_{j_{k}}
    {\cal B}_{j_n}\!\!\cdots\!{\cal B}_{j_{k+1}}
 \nl&\quad
   \times {\cal B}_{j_{k-1}} \!\!\cdots\!{\cal B}_{j_1} {\cal F}
   -i {\sum_j}'\!{\cal A}_{\bar j} %\big(
   {\cal B}_{j} {\cal B}_{j_n}\!\!\cdots\!
     {\cal B}_{j_1}{\cal F}.  %\big).
 \end{align}
 The sum $\sum'$ runs over all $j\neq j_k; k=1,\cdots,n$.
 Denote
 \bsube \label{calFpm}
 \be \label{calFp}
   {\cal F}_{{\bf j}_k}^{(n-1)}
 \equiv
   {\cal B}_{j_n}\!\!\cdots\!{\cal B}_{j_{k+1}}{\cal B}_{j_{k-1}}
   \!\!\cdots\!{\cal B}_{j_1} {\cal F},
 \ee
 and
 \be \label{calFm}
    {\cal F}_{{\bf j}j}^{(n+1)}
  \equiv  {\cal B}_{j} {\cal B}_{j_n}\cdots
     {\cal B}_{j_1}{\cal F}.
 \ee
 \esube
 Equation (\ref{dotFn0}) can now be recast as
 \begin{align} \label{dotFn}
    \partial_t {\cal F}_{\bf j}^{(n)}
  &=
   \ti{\cal F}_{\bf j}^{(n)}
    -i \sum_{k=1}^{n} (-)^{n-k} \ti{\cal C}_{j_{k}}
     {\cal F}_{{\bf j}_k}^{(n-1)}
 \nl&\quad
    -i {\sum_j}' {\cal A}_{\bar j}{\cal F}_{{\bf j}j}^{(n+1)}.
 \end{align}
   Apparently, an odd-tier AIF is a Grassmann variable, while
 an even-tier one an ordinary c-number.
 This is the Grassmann parity that
 will affect the explicit expressions of
 the ${\cal A}$- and ${\cal C}$-terms in
 \Eq{dotFn} in PI repression, and consequently
 their ADO counterparts in operator level; see \Eqs{calCA}.

\subsection{Auxiliary density operators dynamics}
\label{theom0B}

  The $n^{\rm th}$-order ADO
 $\rho^{(n)}_{\bf j}$  can now be defined via
 the AIF ${\cal F}^{(n)}_{\bf j}$ as [cf.\ \Eq{U1_def}]
 \bsube \label{rhon_def}
\be \label{rhon_bfj}
   \rho^{(n)}_{\bf j}(t)\equiv {\cal U}^{(n)}_{\bf j}(t,t_0)\rho(t_0),
\ee
 with %the path-integral expression of the propagator of
 \be \label{Un_bfj}
   {\cal U}^{(n)}_{\bf j}
  (\bm\psi,t;\bm\psi_0,t_0)
 \equiv\! \int_{\bm\psi_0[t_0]}^{\bm\psi[t]}
   \!\!  {\cal D}{\bm\psi}
     e^{iS[\psi]} {\cal F}^{(n)}_{\bf j}[\bm\psi] e^{-iS[\psi']} .
\ee
\esube
 Similarly, the auxiliary operators
 $\ti{\rho}^{(n)}_{\bf j}$, ${\rho}^{(n-1)}_{{\bf j}_k}$,
 and ${\rho}^{(n+1)}_{{\bf j}j}$
 are defined via $\ti{\cal F}^{(n)}_{\bf j}$,
 ${\cal F}^{(n-1)}_{{\bf j}_k}$,
 and ${\cal F}^{(n+1)}_{{\bf j}j}$, respectively.
 The leading term in the $n^{\rm th}$-tier
 AIF such as ${\cal F}^{(n)}$ or $\ti{\cal F}^{(n)}$
 is of the $(2n)^{\rm th}$-order in the system--bath
 coupling; so is that
 in each individual $\rho^{(n)}_{\bf j}$  or $\ti{\rho}^{(n)}_{\bf j}$.

   The EOM for $\rho^{(n)}$ can then be
 readily obtained via \Eq{dotFn} and (\ref{rhon_def}).
 Together with \Eq{dotrho0} and setting
  $\rho^{(0)}\equiv \rho$, we have
\bsube\label{dotrhon}
 \begin{align}
   \dot\rho
  &=
   -i{\cal L}\rho -i\sum_{j}{\cal A}_{\bar j}\rho^{(1)}_{j},
 \label{dotrhonA} \\
%%%
  \dot\rho^{(n)}_{\bf j}
  &=
   -i{\cal L}\rho^{(n)}_{\bf j} + \ti\rho^{(n)}_{\bf j}
   -i \sum_{k=1}^{n}(-)^{n-k} \ti{\cal C}_{j_k}\rho^{(n-1)}_{{\bf j}_k}
 \nl &\quad
   -i{\sum_{j}}'\!{\cal A}_{\bar j}\rho^{(n+1)}_{{\bf j}j};
 \qquad\qquad\qquad   n>0.
 \label{dotrhonB}
\end{align}
\esube
 Here, ${\cal A}_{\bar j}\equiv {\cal A}_{\mu}^{\bar\sigma}$
 and $\ti{\cal C}_j\equiv \ti{\cal C}_{\alpha\mu}^{\sigma}$
 are superoperators. Their PI expressions
 that follow the Grassmann algebra
 are given in \Eqs{calAs} and (\ref{calC_path0}), respectively.
 As the Grassmann parity associated with AIFs,
 the actions of these superoperators in \Eq{dotrhon}
 are given as
\bsube \label{calCA}
 \be\label{calA}
  {\cal A}_{\bar j}{\rho}^{(m)}
  = a_{\mu}^{\bar\sigma}{\rho}^{(m)}
   +(-)^{m}{\rho}^{(m)}a_{\mu}^{\bar\sigma},
 \ee
 and
 \be\label{calC}
   \ti{\cal C}_{j}\rho^{(m)}
 = \sum_{\nu}\big\{C^{0,\sigma}_{\alpha\mu\nu}
     a^{\sigma}_{\nu}\rho^{(m)}
    -(-)^{m}C^{0,\bar\sigma}_{\alpha\nu\mu}
     {\rho}^{(m)}a^{\sigma}_{\nu}\big\}.
 \ee
\esube

  The $n$ indexes in $\rho^{(n)}_{\bf j}\equiv
 \rho^{(n)}_{j_1j_2\cdots j_n}$
 should all be distinct, due to
 the Grassmann or Fermion anticommutation relation
 associated with the index swap.
 Recall that $j\equiv \{\alpha\mu\sigma\}$.
  As results, $\rho^{(n)}=0$,
 if $n>2N_{\alpha}N_c$. Here,
 $N_{\alpha}$ denotes the number of electrodes;
 $N_c$ is the number of spin-orbitals that couple
 directly to the electrodes, [i.e,
 those of $C_{\alpha\mu\mu}(t,\tau)\neq 0$].
 The factor 2 arises from the two possible signs
 of $\sigma=+$ and $-$.

  The EOM formalism presented in \Eq{dotrhon} is finite
 but yet to be closed due to the fact that
 $\{\ti\rho^{(n)}_{\bf j}\}$, as
 defined via $\ti{\cal F}^{(n)}_{\bf j}$
 [\Eq{tiAIF}], involves both the ${\cal B}$-
 and $\ti{\cal B}$-types functionals.
  The desired hierarchy should be constructed
 only via the ${\cal B}$-type functionals.
   To obtain a closed HEOM formalism,
 it requires that $\ti{\cal B}$-type
 functionals [\Eq{ticalBs}] be expressed
 in terms of the ${\cal B}$'s [\Eq{calBs}].
  This can be achieved via the
 extended Meier-Tannor parameterization
 method,\cite{Xu07031107,Jin07134113,Yan05187,Xu029196,Mei993365}
  in which correlation functions
 $C^{\sigma}_{\alpha\mu\nu}(t)$ are expanded in an
 exponential series.
 This method was adopted in our previous
 development of QDT formalism for
 bosonic bath cases, in which the path integral
 variables are all c-numbers.\cite{Xu07031107,Jin07134113}
  We shall treat the parametrization approach
 to HEOM for Fermion reservoirs
 later; see \Sec{ththeo2} and \App{thapp_theo2}.

\subsection{The multiple-frequency dispersed hierarchy construction}
\label{theom0C}

  Proposed here is the MFD  hierarchy scheme, in which
 \be\label{rho_phi}
   \rho^{(n)}_{\bf j}(t)
 = \int d{\omega_1}\cdots\int d{\omega_n}
  \phi^{(n)}_{j_1\cdots j_n}(\omega_1,\!\cdots\!,\omega_n;t).
 \ee
 The desired MFD-HEOM [cf.\ \Eq{final_theo}]
 will be constructed for a proper
 \be\label{phin}
   \phi^{(n)}_{\bf j}({\bm\omega};t)\equiv
   \phi^{(n)}_{j_1\cdots j_n}(\omega_1,\!\cdots\!,\omega_n;t).
 \ee

  Let us start with the time--independent chemical potential case,
 in which the reservoir correlation functions is stationary,
 $C^{\sigma}_{\alpha\mu\nu}(t,\tau)=C^{\sigma}_{\alpha\mu\nu}(t-\tau)$,
 and can be expressed in terms of
 spectrum functions as \Eq{Gam_def}.
 In this case, \Eq{calBs0} can be recast as
 \be \label{hatB}
   B_{j}(t;\{\psi\}) =
    \int d\omega
    \hat B_{j}(\omega,t;\{\psi\}),
 \ee
 with
 \be \label{hatBw}
  \hat B_j(\omega,t;\{\psi\})
 = \!\sum_{\nu}\!
  \int_{t_0}^{t}\!d\tau e^{i\sigma \omega(t-\tau)}
    \Gamma^{\sigma}_{\alpha\mu\nu}\!(\omega)a^{\sigma}_{\nu}[\psi(\tau)].
 \ee
 The frequency--dispersed $\hat B'_j$ is
 similar to $\hat B_j$, except that the
 $\Gamma^{\sigma}_{\alpha\mu\nu}$ oh the rhs of \Eq{hatBw} is replaced by
 $\Gamma^{\bar\sigma\,\ast}_{\alpha\mu\nu}=\Gamma^{\bar\sigma}_{\alpha\nu\mu}$
 [cf.\ the first identity of \Eq{cwsym}].
    As results, \Eq{calBs} can be expressed as
 \be \label{B2Bw}
   {\cal B}_{j} \equiv {\cal B}^{\sigma}_{\alpha\mu}(t;\{{\bm\psi}\})
  = \int d\omega \hat{\cal B}_{j}(\omega,t;\{{\bm\psi}\}) ,
 \ee
 with $\hat{\cal B}_{j} \equiv -i\big(\hat B_{j} - \hat B'_{j}\big)$.

   In the presence of time-dependent chemical potentials
 applied on electrodes, the correlation functions
 include also the nonstationary phase factors \
 [cf.\ \Eq{corr2}]. The frequency dispersed
 $\hat{\cal B}_{j}$--functional reads now
 \begin{align}\label{hatcalB}
  \hat{\cal B}_{j}
  &= -i \sum_{\nu}\int_{t_0}^{t}\!d\tau\,
       e^{i\sigma\omega(t-\tau)}
   \exp\left[ \sigma i\! \int_\tau^t\!dt'\Delta_\alpha({t'})
     \right]
 \nla \times
  \left\{
    \Gamma^{\sigma}_{\alpha\mu\nu}(\omega)a^{\sigma}_{\nu}[\psi(\tau)]
   -\Gamma^{\bar\sigma}_{\alpha\nu\mu}(\omega)a^{\sigma}_{\nu}[\psi'(\tau)]
  \right\}.
 \end{align}
 It satisfies
 \be\label{dothatB}
   \partial_t \hat{\cal B}_j =
   i\sigma [\omega+\Delta_{\alpha}(t)] \hat{\cal B}_j
    + {\cal C}_{j}(\omega),
 \ee
 with ${\cal C}_{j}(\omega)={\cal C}_{j}(\omega;\{\bm\psi[t]\})$ being
 \be\label{calCE0}
   {\cal C}_{j}
  = \sum_{\nu}\left\{ \Gamma^{\sigma}_{\alpha\mu\nu}(\omega)
    a^{\sigma}_{\nu}[\psi(t)]
 -\Gamma^{\bar\sigma}_{\alpha\nu\mu}(\omega)
   a^{\sigma}_{\nu}[\psi'(t)] \right\}.
 \ee
%%%

  The $n^{\rm th}$-tier AIFs, ${\cal F}^{(n)}_{\bf j}$ of \Eq{AIF},
 can now be dispersed as
 \be \label{hatFn}
  {\cal F}^{(n)}_{\bf j} =
   \int d{\bm\omega} \hat{\cal F}^{(n)}_{\bf j}({\bm\omega};t;\{\bm\psi\}),
 \ee
 where [noting that
   $\hat{\cal B}_{j_k}\equiv \hat{\cal B}_{j_k}(\omega_k,t;\{\bm\psi\})$]
 \be\label{AIFw}
  \hat{\cal F}^{(n)}_{\bf j}({\bm\omega};t;\{\bm\psi\})
 \equiv
   \hat{\cal B}_{j_n}\!\!\cdots\hat{\cal B}_{j_2}\hat{\cal B}_{j_1}{\cal F}.
 \ee
 The proper MFD-ADO, $\phi^{(n)}_{\bf j}$ in \Eq{phin},
 for closing the hierarchy is now determined as
  \bsube \label{phihatUall}
  \be \label{rhosfnw_def}
  \phi^{(n)}_{\bf j}({\bm\omega};t)=
  \hat{\cal U}^{(n)}_{\bf j}({\bm\omega};t,t_0)\rho(t_0),
 \ee
 with
 \be \label{calGPInw_def}
   \hat{\cal U}^{(n)}_{\bf j}
 = \int_{{\bm\psi}_0[t_0]}^{{\bm\psi}[t]}
  \!\!  {\cal D}{\bm\psi}
   \, e^{iS[\psi]} \hat{\cal F}^{(n)}_{\bf j} e^{-iS[\psi']} .
 \ee
 \esube
 Note that the Grassmann anticommutation relation is now manifested
 not only in the $j$-indexes, but also in the
 associated frequencies; cf.\ \Eq{AIFw}.

 The time-derivative on \Eq{AIFw} can be obtained as
 \begin{align} \label{dotcalFn}
   \partial_t \hat{\cal F}^{(n)}_{\bf j}
 &=
  i\Omega^{(n)}_{\bm\alpha\bm\sigma}\hat{\cal F}^{(n)}_{\bf j}
     -i \sum_{k=1}^{n} (-)^{n-k}{\cal C}_{j_k}(\omega_k)
    \hat {\cal F}^{(n-1)}_{{\bf j}_k}
 \nl&\quad
  -i \int d\omega {\sum_j} {\cal A}_{\bar j}
   \hat{\cal F}^{(n+1)}_{{\bf j}j}({\bm\omega},\omega,t;\{\bm\psi\}),
 \end{align}
%%%
 with $
   \Omega^{(n)}_{\bm\alpha\bm\sigma}
  = \sum_k\sigma_k [\omega_k+\Delta_{\alpha_k}(t)]$.
 The MFD-HEOM for $\phi^{(n)}_{\bf j}$
  can then be readily obtained.

\section{Hierarchical equations of motion with multiple frequency
  dispersion}
\label{ththeo}

  The final MFD-HEOM  formalism at operator level
 reads as
 (setting $\phi^{(-1)}\equiv \Omega^{(0)}\equiv 0$
   and $\phi^{(0)}\equiv \rho$)
%%%%%
 \begin{align} \label{final_theo}
   \dot\phi^{(n)}_{\bf j}({\bm\omega},t)
 &=
   -i\big[{\cal L}-\Omega^{(n)}_{\bm\alpha\bm\sigma}(\bm\omega,t)\big]
    \phi^{(n)}_{\bf j}({\bm\omega},t)
 \nl&\quad
   -i\sum_{k=1}^n (-)^{n-k}
   {\cal C}_{j_k}(\omega_k)
   \phi^{(n-1)}_{{\bf j}_k}({\bm\omega}'_k,t)
 \nl&\quad
   -i\int d\omega \sum_{j\equiv\{\alpha\mu\sigma\}}\!\!\!\!
   \ {\cal A}^{\bar\sigma}_{\mu}
  \phi^{(n+1)}_{{\bf j}j}({\bm\omega},\omega,t).
 \end{align}
%%%%%
Here, ${\bm\omega}'_k\equiv
 \{\omega_1,\!\!\cdots\!\!,\omega_{k-1},\omega_{k+1},\!\!\cdots\!\!,
  \omega_n\}$,
%%%
\be \label{Omgn_tran}
   \Omega^{(n)}_{\bm\alpha\bm\sigma}(\bm\omega,t)
  = \sum_{k=1}^{n}\sigma_k [\omega_k+\Delta_{\alpha_k}(t)],
 \ee
 and [cf.\ \Eqs{calCA} and (\ref{calCE0})]
 \bsube \label{CAphi}
 \begin{align}
 {\cal A}^{\bar\sigma}_{\mu}\phi^{(m)}
&=
    a^{\bar\sigma}_{\mu}\phi^{(m)}
  +(-)^{m}\phi^{(m)}a^{\bar\sigma}_{\mu},
 \label{CAphiA}\\
   {\cal C}_{j}(\omega)\phi^{(m)}
 &=
  \sum_\nu\big[\Gamma^{\sigma}_{\alpha\mu\nu}(\omega)
    a^\sigma_\nu\phi^{(m)}
 \nl&\quad\quad
   -(-)^{m}\Gamma^{\bar\sigma}_{\alpha\nu\mu}(\omega)\phi^{(m)}
   a^\sigma_\nu\big].
 \label{CAphiB}
 \end{align}
\esube
  The initial conditions to \Eq{final_theo}
 are in fact those of steady--state solutions;
 see \Sec{thsum} for details.
  The transient current [\Eq{currI}] is evaluated
 via the first--tier frequency--dispersed ADO,
 $\phi^{(1)}_j(\omega,t)\equiv\phi^{\sigma}_{\alpha\mu}(\omega,t)$, as
 \be\label{currIw}
   I_{\alpha}(t) = -2\,{\rm Im} \int\!d\omega \sum_{\mu} {\rm tr}_{\rm s}
    \left\{a_{\mu}\phi^{+}_{\alpha\mu}(\omega,t)\right\}.
 \ee

 As inferred from \Eq{hatcalB} and
 the first identity of \Eq{cwsym},
 we have (denoting ${\bar j}\equiv \{\alpha\mu{\bar\sigma}\}$)
 \be\label{rhonsym}
  [\phi^{(n)}_{j_1\cdots j_n}]^{\dg}
   = \phi^{(n)}_{{\bar j}_n\cdots {\bar j}_1}
   = (-)^{[\frac{n}{2}]} \phi^{(n)}_{{\bar j}_1\cdots {\bar j}_n},
 \ee
 which can be abbreviated as $\big[\phi^{(n)}_{\bf j}\big]^{\dg} =
 (-)^{[\frac{n}{2}]} \phi^{(n)}_{\bar{\bf j}}$, with
 $[\frac{n}{2}]\equiv {\rm Int}(n/2)$ being for
 the number of index-swaps involved.
 This is the generalization
 of \Eq{rho1sym}. The Grassmann anticommutation relation
 in $\phi^{(n)}_{\bf j}(\bm\omega;t)$ involves
 not only the $j$-indexes, but also the
 associated frequencies, such as
 $\phi_{j'j}(\omega',\omega;t)=-\phi_{jj'}(\omega,\omega';t)$.
 As results, $\phi_{jj}\neq 0$, although its
 double--frequency integral vanishes.

 The MFD--HEOM formalism is exact,
 but with an infinite hierarchy.
 The truncation is however rather
 trivial,\cite{Xu07031107} as the leading term in
 an $n^{\rm th}$--tier ADO
 $\phi^{(n)}$ is of the $(2n)^{\rm th}$--order system--reservoir
 coupling [cf.\ \Eq{AIFw}].
 In practice, one can set all $\phi^{(n>N_{\rm trun})}\approx 0$,
 followed by the convergence test with increasing $N_{\rm trun}$
 for the anchor tier of the hierarchy.
 We will show in \Sec{thtier2} that
 the present theory with $N_{\rm trun}=2$
 recovers the real--time diagrammatic formalism
 developed by Sch\"{o}n and
 coworkers.\cite{Sch9418436,Sch94423,Kon9531,Kon9616820,Kon961715}
 Also, we will see soon that in the absence
 of Coulomb interaction, the Grassmann parity
 of ${\cal A}^{\bar\sigma}_{\mu}$ defined in \Eq{CAphiA}
 leads to $\phi^{(3)}=0$ without approximation;
 the present theory is exact of
 $n_{\rm max}=2$ for single--particle systems.

\section{Equations of motion for single-particle Hamiltonian systems}
\label{thgreen}

\subsection{Auxiliary single-particle density matrices }
\label{thgreenA}

  Consider a single-particle system, described by
 \be \label{H1e}
   H = \sum_{\mu\nu} h_{\mu\nu} a^{\dg}_{\mu}a_{\nu}.
 \ee
 The electronic dynamics in such a system can
 be characterized by
 {\it reduced single-particle density matrix} (RSPDM),
 ${\bm \varrho}$, defined via its elements of
 \be \label{DM0}
    \varrho_{\mu\nu}(t) \equiv {\rm tr}_{\rm s}
     \big[a^{\dg}_{\nu}a_{\mu}\rho(t)].
 \ee

  Presented in \App{thapp_green} is the
 derivation of RSPDM dynamics via MFD-HEOM,
 which is of $n_{\rm max}=2$ for the single-particle
  system without approximation.
  This results from the Grassmann parity
  associated ${\cal A}^{\bar\sigma}_{\mu}$
 [\Eq{CAphiA}], whose action on single-particle
  operators via \Eq{final_theo} will terminate at $n_{\rm max}=2$.
 Only the first- and the second-tier auxiliary
 density matrices,
 $\bm\varphi_{\alpha}$ and $\bm\varphi_{\alpha'\alpha}$,
 are required for a single-particle system.
 These auxiliary frequency-resolved density matrices
 can be defined via their elements of
 \be \label{DM1}
     [\varphi_{\alpha}(\omega,t)]_{\mu\nu}
 \equiv {\rm tr}_{\rm s}
    [a_{\mu}\phi^{+}_{\alpha\nu}(\omega,t)],
 \ee
 and
 \be \label{DM2}
   [\varphi_{\alpha'\alpha}(\omega',\omega,t)]_{\mu\nu}
  \equiv {\rm tr}_{\rm s}
    [\phi^{+,-}_{\alpha\nu,\alpha'\mu}(\omega,\omega',t)].
 \ee
 It can be shown via \Eq{rhonsym} that
 \be \label{vphi2_sym}
    [\bm\varphi_{\alpha'\alpha}(\omega',\omega,t)]^{\dg}
  = \bm\varphi_{\alpha\alpha'}(\omega,\omega',t).
 \ee
%%%%%%%%%%%%%%%%%%
   The final results are summarized as follows.
 \bsube \label{final_green}
 \begin{align}
    i\dot{\bm\varrho}
 &=
    [{\bm h},\bm\varrho] - \sum_{\alpha}\int\!\! d\omega
      \big[\bm\varphi_{\alpha}(\omega,t)
       - \bm\varphi^{\dg}_{\alpha}(\omega,t) \big],
 \label{final_green0} \\
     i\dot{\bm\varphi_{\alpha}}
 &=
   [{\bm h}-\omega-\Delta_{\alpha}(t)]\bm\varphi_{\alpha}
    +{\bm S}_{\alpha}(\omega) -\bm\varrho {\bm J}_{\alpha}(\omega)
 \nl&\quad
    +\sum_{\alpha'}\int d\omega'
     \bm\varphi_{\alpha'\alpha}(\omega',\omega,t),
 \label{final_green1} \\
   i\dot{\bm\varphi}_{\alpha'\alpha}
 &=
     -[\omega+\Delta_{\alpha}(t)-\omega'-\Delta_{\alpha'}(t)]
       \bm\varphi_{\alpha'\alpha}
 \nl&\quad
     + {\bm J}_{\alpha'}(\omega')\bm\varphi_{\alpha}(\omega,t)
     - \bm\varphi^{\dg}_{\alpha'}(\omega',t){\bm J}_{\alpha}(\omega).
 \label{final_green2}
 \end{align}
 \esube
%%%
 Here,
 \be\label{S_def}
  {\bm S}_{\alpha}(\omega)\equiv f_{\alpha}(\omega){\bm J}_{\alpha}(\omega)
  = (\bm\Gamma^+_{\alpha})^{T},
 \ee
 with $f_{\alpha}(\omega)$
  being the Fermi distribution function,
 and ${\bm J}_{\alpha}(\omega)={\bm J}^{\dg}_{\alpha}(\omega)$ [\Eq{Jw_def}]
 the spectral density function matrix of the reservoir $\alpha$.
 The last identity in \Eq{S_def}
 states that ${\bm S}_{\alpha}$ is equivalent to
 the $\bm\Gamma^{+}_{\alpha}$--matrix transpose.
 Note that \Eq{final_green} can be recast in the
 standard form of linearly coupled equations,
 see \Eq{app_final_green}.

  The transient current of \Eq{currI} from the $\alpha$-lead
 to the system can be expressed as
 \be\label{green_I}
  I_{\alpha}(t) = -2\,{\rm Im}\int d\omega\,
     {\rm tr}\big[\bm\varphi_{\alpha}(\omega,t) \big].
 \ee
 The matrix trace is used here.
  Together with \Eq{final_green1}, the above expression
 leads to ${\rm tr}\dot{\bm\varrho}(t)=\sum_\alpha I_{\alpha}(t)$.
 This is the flux conservation relation.
 %%%%%%%%%%%%%%%%%%%%%%%%

 \subsection{Steady-state current and
     the Landauer-B\"{u}ttiker formula}
 \label{thgreenB}
   Consider now the stationary solutions,
 $\bm\varrho = \bm\varrho(t\rightarrow \infty)$ and
 $\bm\varphi(\omega)\equiv \bm\varphi(\omega,t\rightarrow \infty)$,
 to \Eq{final_green}
 in the absence of time--dependent external bias potentials.
  The steady--state solution to \Eq{final_green2}
 is
 \be\label{green2C}
   \bm\varphi_{\alpha'\alpha}(\omega',\omega)
 = \frac{
    {\bm J}_{\alpha'}(\omega') \bm\varphi_{\alpha}(\omega)
   -\bm\varphi_{\alpha'}^\dg(\omega'){\bm J}_\alpha(\omega)
   } {\omega-\omega'+i0^+}.
 \ee

  To proceed, let $x(\omega)=\sum_\alpha x_{\alpha}(\omega)$;
 where $x \in \{{\bm J}, {\bm S}, \bm\varphi\}$.
  Adopt also the common definitions
  of retarded self--energy ${\bm\Sigma}$
  and Green's function ${\bm G}$ as
 \bsube \label{selfEG}
 \be \label{selfE}
   \bm\Sigma(\omega)
    \equiv \int d\omega'\frac{{\bm J}(\omega')}{\omega-\omega'+i0^+}\, ,
 \ee
 and
 \be \label{greenG}
  {\bm G}(\omega)
   \equiv [\omega-{\bm h}-\bm\Sigma(\omega)]^{-1}.
 \ee
 \esube
 Substituting \Eqs{green2C} and (\ref{selfEG})
 into \Eq{final_green1} in steady state leads to
 \begin{align} \label{green1a}
    {\bm G}^{-1}(\omega)\bm\varphi_{\alpha}(\omega)
 &=
    {\bm S}_{\alpha}(\omega)
  -\bm\varrho {\bm J}_{\alpha}(\omega)
 \nla
    - \int d\omega'\frac{
    {\bm\varphi}^\dg(\omega')}{\omega-\omega'+i0^+}{\bm J}_\alpha(\omega).
 \end{align}
 Summing over $\alpha$ in the above equation,
 followed by some simple algebra,
 will obtain the relation of
 \be \label{green1}
   \int d\omega'\frac{
    \bm\varphi^\dg(\omega')}{\omega-\omega'+i0^+}
 = {\bm S}{\bm J}^{-1}-\bm\varrho
  -{\bm G}^{-1}\bm\varphi{\bm J}^{-1}.
 \ee
 Therefore
 \be\label{phialphi}
  \bm\varphi_{\alpha}
  ={\bm G}({\bm S}_{\alpha}-
   {\bm S}{\bm J}^{-1}{\bm J}_{\alpha})
     +\bm\varphi{\bm J}^{-1}{\bm J}_{\alpha}.
 \ee

  For electrodes of similarity,
 ${\bm J}_{\alpha}(\omega)\sim {\bm J}_{\rm \alpha'}(\omega)$,
 up to trivial constants,
 the last term in \Eq{phialphi}
 does not contribute to current of \Eq{green_I},
 as it satisfies
\[
 {\rm Im}\!\int\!d\omega\,{\rm tr}(\bm\varphi{\bm J}^{-1}{\bm J}_{\alpha})
 \propto {\rm Im}\!\int\!d\omega\,{\rm tr}\,\bm\varphi = 0.
\]
 Thus, the steady--state current assumes
 \be\label{Ialpha}
 I_\alpha=-2\,{\rm Im}\!\int\!d\omega\,
 {\rm tr}\big[{\bm G}({\bm S}_{\alpha}-
  {\bm S}{\bm J}^{-1}{\bm J}_{\alpha})\big].
 \ee

   Consider now the current contribution from lead L
 in a two--terminal setup.
 The quantity in the parentheses in the rhs
 of \Eq{phialphi} can be evaluated as
 \be\label{SLSJ}
  {\bm S}_{\rm L}-{\bm S}{\bm J}^{-1}{\bm J}_{\rm L}
  = [f_{\rm L}(\omega)-f_{\rm R}(\omega)]
   \bar{\bm J}(\omega) ,
 \ee
 with
 \be\label{Jbar}
   [\bar{\bm J}(\omega)]^{-1}
  =[{\bm J}_{\rm L}(\omega)]^{-1}
    + [{\bm J}_{\rm R}(\omega)]^{-1}.
 \ee
 The identity  ${\bm A}({\bm A}+{\bm B})^{-1}{\bm B}
  =({\bm A}^{-1}+{\bm B}^{-1})^{-1}$ is used here.
 Thus \Eq{Ialpha} for L--lead reads
 \be \label{I_st}
   I_{\rm L}=-2\,{\rm Im}\!\int\!d\omega\,
    {\rm tr}[(f_{\rm L}-f_{\rm R}){\bm G}\bar{\bm J}\,].
 \ee
 This is exactly the Landauer--B\"{u}ttiker's
 result,\cite{Dat95, Mei922512, Lan92110,But88317}
 with the original factor of $(2\pi)^{-1}$ being included
 now in the reservoir spectral density matrix.

\section{Relation to the real--time diagrammatic formalism}
\label{thtier2}

  In this section we shall show that
 the real--time diagrammatic formalism developed
 by Sch\"{o}n and co-workers\cite{Sch9418436,Sch94423,%
 Kon9531,Kon9616820,Kon961715} corresponds to a
 second--tier approximation,
 i.e., setting all $\phi^{(n\geq 3)}=0$,
 within the framework of the
 present MFD--HEOM of \Eq{final_theo}.
 For clarity, we choose for demonstration
 the same system
 as Ref.\,\onlinecite{Kon9616820},
 where Kondo physics was studied
 via steady--state two--terminal current measurement
 in the strong Coulomb interaction regime.
 The Anderson impurity Hamiltonian is adopted for the quantum dot system,
 \be\label{andersonH}
   H= \sum_{\mu}\epsilon_{\mu}\hat n_{\mu}
     + U \sum_{\mu<\mu'}\hat n_{\mu}\hat n_{\mu'}.
 \ee
 Here, $\hat n_{\mu} = a^{\dg}_{\mu}a_{\mu}$, with
 $\mu$ being the spin index; thus
 $C^{\sigma}_{\alpha\mu\nu}(t)=0$ if $\mu\neq \nu$,
 and
  $\Gamma^{\sigma}_{\alpha\mu\nu}(\omega)
   =\Gamma^{\sigma}_{\alpha\mu}(\omega)\delta_{\mu\nu}$.
%%%
  In the strong Coulomb interaction regime,
 $a_{\mu}=|0\ra\la\mu|$, as the basis set
 involves only the zero and single occupation dot states
 and can be denoted as $\{|s\ra; s=0, \mu\}$.
  As results, $\rho\equiv \phi^{(0)}$
 and its associated ADOs $\phi^{(n>0)}_{\bf j}$ in \Eq{final_theo}
 are all $(m+1)\times (m+1)$ matrices,
 with $m$ being the number of spin states.
   The measured current $I= I_{\rm L}-I_{\rm R}$ is then of
 [cf.\ \Eq{currIw}]
 \be \label{currIdot}
    I_{\alpha} =-2\, {\rm Im} \int\!d\omega \sum_{\mu}
   [\phi^{+}_{\alpha\mu}(\omega)]_{\mu 0},
 \ee
 with $[\phi^{\sigma}_{{\alpha}\mu}(\omega)]_{s s'}\equiv
 \la s|\phi^{\sigma}_{{\alpha}\mu}(\omega)| s'\ra$
 being the matrix element of the first--tier ADO.
 Note also that the reduced
 density matrix in a steady--state is diagonal,
 $\rho^{\rm st}_{ss'}=\delta_{ss'}\rho^{\rm st}_{ss}
 \equiv \delta_{ss'}P_s$, with $\sum_s P_s=P_0+\sum_{\mu}P_{\mu}=1$,
 as the density matrix in the dot-state representation
 is normalized.
 Under the steady-state condition,
 \Eq{final_theo} with $n=0$ reads
 \be \label{trace_rho}
   0=\dot P_{\mu} =-2\,{\rm Im}\sum_\alpha\!\!\int \!\!d\omega
    \big[\phi^{+}_{\alpha\mu}(\omega)\big]_{\mu0},
 \ee
 while it for the first--tier auxiliary density matrix element,
 $\la\mu|\dot\phi^{+}_{\alpha\mu}(\omega)|0\ra=0$,
 reads
 \begin{align}\label{anderson1}
   0
 &=
   (\epsilon_\mu-\omega-i0^+)
   \big[\phi^{+}_{\alpha\mu}(\omega)\big]_{\mu0}
  +\Gamma^{+}_{\alpha\mu}(\omega)P_0
  -\Gamma^{-}_{\alpha\mu}(\omega)P_{\mu}
 \nla
 + \int\!\! d\omega'\Big\{
  \big[\phi^{+-}_{\alpha\mu\mu}(\omega,\omega')\big]_{00}
  +\sum_{\mu'}\big[\phi^{+-}_{\alpha\mu\mu'}(\omega,\omega')\big]_{\mu\mu'}
  \Big\}.
 \end{align}
 Here,
  $\phi^{+-}_{\alpha\mu\mu'}
 \equiv \sum_{\alpha'}\phi^{+,-}_{\alpha\mu,\alpha'\mu'}$
 are the coupling second-tier ADOs.
 The above equation does not involve
 $\phi^{++}_{\alpha\mu\alpha'\mu'}$, as
 $\la \mu|[a_{\mu'},
  \phi^{++}_{\alpha\mu\alpha'\mu'}(\omega,\omega')]|0\ra
  =0$.
  Denote in the following are also
 $\Gamma^{\pm}_{\mu}\equiv\sum_{\alpha}\Gamma^{\pm}_{\alpha\mu}$
 and $\phi^{\pm}_{\mu}\equiv\sum_{\alpha}\phi^{\pm}_{\alpha\mu}$.

  Presented in the following are
 the analytical results under the second--tier approximation,
 in which we set
 the third-tier ADOs $\phi^{(3)}=0$.
 Thus, the second-tier ADOs are determined
 completely by the first-tier ADOs.
 In the steady--state condition, we obtain
%%%%%%%%%%%%
 \begin{align*}
   \big[\phi^{+-}_{\alpha\mu\mu}\big]_{00}
  &\approx\frac{\Gamma^{-}_{\mu}(\omega')
   \big[\phi^{+}_{\alpha\mu}(\omega)\big]_{\mu0}
   -\Gamma^{-}_{\alpha\mu}(\omega)
   \big[\phi^{+}_{\mu}(\omega')\big]^\ast_{\mu0}}
   {\omega-\omega'+i0^+},
   \\
  \big[\phi^{+-}_{\alpha\mu\mu'}\big]_{\mu\mu'}
 &\approx
  \frac{\Gamma^{+}_{\mu'}(\omega')
    \big[\phi^{+}_{\alpha\mu}(\omega)\big]_{\mu0}
  -\Gamma^{+}_{\alpha\mu}(\omega)
  \big[\phi^{+}_{\mu'}(\omega')\big]^\ast_{\mu' 0}}
  {\omega-\omega'-\epsilon_\mu+\epsilon_{\mu'}+i0^+}.
 \end{align*}
 Substituted into \Eq{anderson1} leads to
 \begin{align}\label{phi1s}
  \big[\phi^{+}_{\alpha\mu}(\omega)\big]_{\mu0}
&=
  \Pi_{\mu}(\omega)\Bigl[
   \Gamma^{+}_{\alpha\mu}(\omega)P_{0}
   -\Gamma^{-}_{\alpha\mu}(\omega)P_{\mu}
 \nl&\quad
  -\Gamma^-_{\alpha\mu}(\omega)\!\!\int\!\! d\omega'
   \frac{\big[\phi^{+}_{\mu}(\omega')\big]^\ast_{\mu 0}}
    {\omega-\omega'+i0^+}
   -\Gamma^+_{\alpha\mu}(\omega)
\nl&\qquad\times
  \sum_{\mu'}\!\!\int\!\! d\omega'
   \frac{\big[\phi^{+}_{\mu'}(\omega')\big]^\ast_{\mu' 0}}
   {\omega-\omega'-\omega_{\mu\mu'}+i0^+}
 \Bigr].
 \end{align}
 Here $\omega_{\mu\mu'}\equiv \epsilon_\mu-\epsilon_{\mu'}$,
 and
 \be\label{Pi_realtime}
   \Pi_{\mu}(\omega)=\frac{1}{\omega-\epsilon_\mu-\Sigma_{\mu}(\omega)+i0^+},
 \ee
 with
 \be\label{Sig_realtime}
   \Sigma_{\mu}(\omega)
 \equiv
   \int\!\! d\omega'
    \Bigl[\frac{\Gamma^-_{\mu}(\omega')}{\omega-\omega'+i0^+}
  + \sum_{\mu'}\!\frac{\Gamma^+_{\mu'}(\omega')}
   {\omega-\omega'-\omega_{\mu\mu'}+i0^+}\Bigr].
 \ee
%%%%%%%%%%%%%%%%%%%%%

  To make connections to the real--time diagrammatic
formulation presented in Ref.\,\onlinecite{Kon9616820},
 we denote
 \be\label{phi1sd}
  \big[\phi^{+}_{\alpha\mu}(\omega)\big]_{\mu0}
  \equiv \sum_{s} Y^{s}_{\mu 0}(\alpha,\mu,\omega)P_{s}.
 \ee
 Here $Y^{s}_{\mu 0}(\alpha,\mu,\omega)$ is
 identical to the $\phi^{s,\mu}_{s,0}(\alpha,\mu,\omega)$
 in Ref.\,\onlinecite{Kon9616820}.
 which is associated with the tunneling of an electron of
 spin $\mu$ and energy $\omega$,
 from $\alpha$--reservoir to the dot, provided
 the initial state of $\rho_{ss}=P_s$.
 Inserting \Eq{phi1sd} to \Eq{phi1s} leads to
\begin{align}\label{Y_realtime}
 Y^{s}_{\mu 0}
&=\Pi_{\mu}(\omega)
  \Big[ \Gamma^{+}_{\alpha\mu}(\omega)\delta_{s0}
   - \Gamma^{-}_{\alpha\mu}(\omega)\delta_{s\mu}
\nl&\qquad
  - \Gamma^-_{\alpha\mu}(\omega)
 \int\!\! d\omega'
   \frac{[Y^{s}_{\mu 0}(\mu,\omega')]^{\ast}}{\omega-\omega'+i0^+}
\nl&\qquad
 -\Gamma^+_{\alpha\mu}(\omega)
 \sum_{\mu'}\!\!\int\!\! d\omega'
 \frac{[Y^{s}_{\mu' 0}(\mu',\omega')]^{\ast}}
 {\omega-\omega'-\omega_{\mu\mu'}+i0^+} \Big],
\end{align}
 with $Y^{s}_{\mu 0}(\mu,\omega)\equiv \sum_{\alpha'}
 Y^{s}_{\mu 0}(\alpha',\mu,\omega)$ adopted in the rhs.
 Equation (\ref{Y_realtime}) is identical to the key
 results, Eqs.\,(50)$-$(51) in Ref.\,\onlinecite{Kon9616820}.
 We have thus demonstrated
 that the second--tier approximation of the
 present MFD--HEOM [\Eq{final_theo}]
 does recover
 the real--time diagrammatic
 formalism.\cite{Sch9418436,Sch94423,Kon9531,Kon9616820,Kon961715}

\section{Hierarchical equations of motion with parametrization}
\label{ththeo2}
 \subsection{Non-Markovian reservoirs via parameterization}
 \label{ththeo2A}

  In this section, we consider an alternative
 hierarchy construction based on exponential
 series expansion of reservoir correlation functions.
   The resulting HEOM formalism will completely
 be in time domain,
 thus has the advantage in numerical implementation.
  The exponential series expansion of reservoirs
 can be achieved via the extended
 Meier-Tannor parametrization method,%
 \cite{Mei993365,Yan05187,Xu029196}
 which has been used in the construction
 of HEOM for systems coupled with
 bosonic bath.\cite{Xu07031107,Jin07134113}

  The fermionic reservoir parametrization starts
 with the following form of spectral density functions,
 \be \label{Jpara}
    J^{\sigma}_{\alpha\mu\nu}(\omega)
  =\sum_{k=0}^K
   \frac{\big(\Gamma^{\mu\nu}_{\alpha k}
    +\sigma i\bar\Gamma^{\mu\nu}_{\alpha k}\big)
     (W^{\mu\nu}_{\alpha k})^2
   }
   {(\omega +\sigma\Omega^{\mu\nu}_{\alpha k})^2
     +(W^{\mu\nu}_{\alpha k})^2
   }.
 \ee
 Involving parameters are all real and positive
 (except for the Drude term with $k=0$
 and $\Omega^{\mu\nu}_{\alpha 0}\equiv 0$),
 and due to $J^{\sigma}_{\alpha\mu\nu}=J^{\sigma\,\ast}_{\alpha\nu\mu}$,
 they satisfy
\[
 (\Gamma^{\mu\nu}_{\alpha k},W^{\mu\nu}_{\alpha k},
 \Omega^{\mu\nu}_{\alpha k},\bar\Gamma^{\mu\nu}_{\alpha k})
 =(\Gamma^{\nu\mu}_{\alpha k},W^{\nu\mu}_{\alpha k},
 \Omega^{\nu\mu}_{\alpha k},-\bar\Gamma^{\nu\mu}_{\alpha k}).
\]
 The corresponding stationary components of
 correlation functions can then be obtained via the FDT
 [\Eq{FDT}] using the contour integration method.
 We have
 \be\label{corr_para}
   C^{\sigma}_{\alpha\mu\nu}(t)
  = \sum_{k=0}^K
   \eta^{\sigma}_{\alpha\mu\nu k}
    e^{-\gamma^{\sigma}_{\alpha\mu\nu k} t}
  +\sum_{m=1}^M
   \check\eta^{\sigma}_{\alpha\mu\nu m}
   e^{-\check\gamma^{\sigma}_{\alpha\!m} t} .
  \ee
 The first term arises from the poles
 of the spectral density functions, with\cite{Jin07134113}
 \bsube \label{drude1}
 \begin{align}
   \gamma^{\sigma}_{\alpha\mu\nu k}
 &= W^{\mu\nu}_{\alpha k} - \sigma i\Omega^{\mu\nu}_{\alpha k},
 \label{gammaD}\\
   \eta^{\sigma}_{\alpha\mu\nu k}
 &= \frac{\big(\Gamma^{\mu\nu}_{\alpha k}
      +\sigma i\bar\Gamma^{\mu\nu}_{\alpha k}\big)
      W^{\mu\nu}_{\alpha k}
    }
    {1+\exp\big[i\beta_{\alpha}(\gamma^{\sigma}_{\alpha\mu\nu k}
             +\sigma i\mu_\alpha)\big]
    } .
 \label{etaD}
 \end{align}
 \esube
 Note that
  $\eta^{\bar\sigma\,\ast}_{\alpha\mu\nu k}
 = e^{i\beta_{\alpha}(\gamma^{\sigma}_{\alpha\mu\nu k}
             +\sigma i\mu_\alpha)}
   \eta^{\sigma}_{\alpha\mu\nu k}$.
 The second term in \Eq{corr_para}, with $M\rightarrow\infty$ in principle,
 arises from the Matsubara poles. The involving parameters
 are\cite{Jin07134113}
 \bsube \label{matsu1}
 \begin{align}
    \check\gamma^{\sigma}_{\alpha m}
 &=   \beta^{-1}_{\alpha}(2m - 1)\pi %\check{\gamma}_{ m}
    - \sigma i\mu_\alpha,
%  = \check\gamma^{\bar\sigma\,\ast}_{\alpha m} ,
 \label{gamMat} \\
    \check\eta^{\sigma}_{\alpha\mu\nu m}
 &=
   \frac{2}{i\beta_{\alpha}}
    J^{\sigma}_{\alpha\mu\nu}(-i\check{\gamma}^{\sigma}_{\alpha m})
   =-\check\eta^{\bar\sigma\,\ast}_{\alpha\mu\nu m}.
 \label{etaMat}
\end{align}
\esube
 The last identity arises from the symmetry relation
 of Fermion spectral density functions.

  The dissipation functional [\Eq{calR}]
 is now decomposed according to \Eq{corr_para} as
 \be\label{calRpara}
  \mathcal{R}=  i \sum_{\bf k}
     \mathcal{A}^{\bar\sigma}_{\mu}\mathcal{B}_{\bf k}
   + i \sum_{\bf m}
     \mathcal{A}^{\bar\sigma}_{\mu}\check {\cal B}_{\bf m}.
 \ee
 with ${\bf k}\equiv(\alpha\mu\nu\sigma k)$ and
 ${\bf m}\equiv(\alpha\mu\sigma m)$;
 see \App{thapp_theo2} for details.
 Presented there is also the derivation
 of HEOM formalism to be summarized as follows.

\subsection{HEOM via parametrization: Final results}
\label{ththeo2C}

  The final HEOM formalism via parametrization of \Eq{Jpara}
 can be written in the following compact form:
 \be\label{final_para}
   \dot\rho_{\ind} =
   -\big[i{\cal L} + \gamma_\ind(t)\big]\rho_\ind
     +\rho^{\{-\}}_\ind+\rho^{\{+\}}_\ind,
 \ee
 with $\gamma_{\sf 0} = \rho_{\sf -1}\equiv 0$,
 and $\rho_{\sf 0}\equiv \rho$ the primary reduced
 density operator.
 The subscript ${\ind} \equiv
  ({\bf k}_1\cdots{\bf k}_p,{\bf m}_1\cdots{\bf m}_q)$
 denotes an ordered set of indexes,
 with
 \[ {\bf k} =(\alpha\mu\nu\sigma k)
 \ \ \text{and}\ \
  {\bf m}=(\alpha\mu\sigma m),\]
 arising
 from the two distinct components of the
 reservoir correlation functions [cf.\ \Eq{corr_para}
 or (\ref{calRpara})]. Note that
 \be\label{rhon_para}
  \rho_{\ind}\equiv
   \rho^{(n)}_{{\bf k}_1\cdots{\bf k}_p,{\bf m}_1\cdots{\bf m}_q};
 \qquad  p+q = n,
 \ee
 is an $n^{\rm th}$--tier ADO; see \Eq{AIF_para} for its
 associating AIF.
  It satisfies the Hermitian conjugate relation of [cf.\
\Eq{rhonsym}]
 \be\label{rhonsym_para}
  [\rho^{(n)}_{{\bf k}_1\cdots{\bf k}_p,{\bf m}_1\cdots{\bf m}_q}]^{\dg}
   = (-)^{[\frac{n}{2}]}
   \rho^{(n)}_{\bar{\bf k}_1\cdots\bar{\bf k}_p,%
   \bar{\bf m}_1\cdots\bar{\bf m}_q},
 \ee
 where $\bar{\bf k} =(\alpha\mu\nu\bar\sigma k)$ and
  $\bar{\bf m}=(\alpha\mu\bar\sigma m)$.

 The parameter $\gamma_{\ind}$  in \Eq{final_para}
 collects the complex ``damping'' parameters
 of the involving reservoir correlation functions,
 see \Eqs{phase_para} and (\ref{dotcalB_para}).
 It reads
 \begin{align} \label{gamma_n}
   \gamma_\ind(t)
 &= \sum^p_{j=1}
    \big[
     \gamma^{\sigma}_{\alpha\mu\nu k}-\sigma i\Delta_{\alpha}(t)
    \big]_{j; \{\alpha,\mu,\nu,\sigma,k\in{\bf k}\}}
 \nla
  +\sum^q_{l=1}
    \big[
     \check\gamma^{\sigma}_{\alpha m}-\sigma i\Delta_{\alpha}(t)
    \big]_{l;\{\alpha,\sigma,m\in{\bf m}\}}.
 \end{align}
 It contains not only the stationary
 components as \Eq{corr_para},
 but also the nonstationary contributions.
 The latter are described by
 the time--dependent chemical potentials $\Delta_{\alpha}(t)$,
 applied on top of the constant $\mu_{\alpha}$
 on electrodes; see \Eq{corr2}.

 The tier-down term in \Eq{final_para} reads
 \be\label{rhondown_para}
 \rho^{\{-\}}_\ind
 = -i\sum^p_{j=1} (-)^{n-j}
   {\cal C}_{{\bf k}_j}
   \rho_{\ind^-_{j}}-i\sum^q_{l=1} (-)^{q-l}
    \check{\cal C}_{{\bf m}_l}
   \rho_{\check\ind^-_l},
 \ee
 with the $(n-1)^{\rm th}$--tier ADOs of
 \bsube\label{rhodown_para}
 \begin{align}
   \rho_{\ind^-_{j}}
 &\equiv
  \rho^{(n-1)}_{{\bf k}_1\cdots{\bf k}_{j-1}
  {\bf k}_{j+1}\cdots{\bf k}_p,{\bf m}_1\cdots{\bf m}_q},
 \\
   \rho_{\check\ind^-_{l}}
 &\equiv
  \rho^{(n-1)}_{{\bf k}_1\cdots{\bf k}_p,%
   {\bf m}_1\cdots{\bf m}_{l-1}
  {\bf m}_{l+1}\cdots{\bf m}_q}.
 \end{align}
 \esube
 ${\cal C}_{{\bf k}_j}$
 and $ \check{\cal C}_{{\bf m}_l}$ in \Eq{rhondown_para}
 are the Liouville--space operator counterparts of
 \Eq{app_calCPI_para} in PI representation,
 with the Grassmann parity associated actions of
 \bsube \label{calC_para}
 \begin{align} \label{Ckrho}
 {\cal C}_{\bf k}\rho_{\ind^{\!-}}
 &\equiv
 \eta^{\sigma}_{\alpha\mu\nu k}a^{\sigma}_{\nu}\rho_{\ind^{\!-}}
  +(-)^{n}\eta^{\bar\sigma\ast}_{\alpha\mu\nu k}
  \rho_{\ind^{\!-}}a^{\sigma}_{\nu},
 \\
 \label{Cmrho}
 \check{\cal C}_{\bf m}\rho_{\check\ind^{\!-}}
 &\equiv\sum_{\nu}
  \check\eta^{\sigma}_{\alpha\mu\nu m}\big[a^{\sigma}_{\nu}\rho_{\check\ind^{\!-}}
  -(-)^{n}
  \rho_{\check\ind^{\!-}}a^{\sigma}_{\nu}\big].
 \end{align}
\esube

  The tier-up term in \Eq{final_para} arises from
 the contribution of $\partial_t {\cal F}=-{\cal RF}$
 to $\partial_t{\cal F}_{\ind}$.  It is given by
 \be \label{rhonup_para}
  \rho^{\{+\}}_\ind
  = -i \sum_{\bf k}
   (-)^q{\cal A}^{\bar\sigma}_{\mu} \rho_{\ind^+_{\bf k}}
  -i  \sum_{\bf m}
    {\cal A}^{\bar\sigma}_{\mu}
      \rho_{\check\ind^+_{\bf m}},
 \ee
 with the $(n+1)^{\rm th}$--tier of ADOs of [cf.\ \Eq{calFup}]
 \bsube \label{rhoup_para}
 \begin{align}
   \rho_{\ind^+_{\bf k}}
 &\equiv
   \rho^{(n+1)}_{{\bf k}_1\cdots{\bf k}_p{\bf k},{\bf m}_1
   \cdots{\bf m}_q},
 \\
   \rho_{\check\ind^+_{\bf m}}
 &\equiv
   \rho^{(n+1)}_{{\bf k}_1\cdots{\bf k}_p,{\bf m}_1\cdots
   {\bf m}_q{\bf m}}.
 \end{align}
 \esube
 The Grassmann parity associated Liouville--space operator
 ${\cal A}^{\bar\sigma}_{\mu}$ in \Eq{rhonup_para} are
 given by [cf.\ \Eq{calA}]
 \be \label{calA_para}
 {\cal A}^{\bar\sigma}_{\mu}\rho_{\ind^{\!+}}
  = a^{\bar\sigma}_{\mu}\rho_{\ind^{\!+}} -(-)^{n}\rho_{\ind^{\!+}}
    a^{\bar\sigma}_{\mu}.
 \ee

 The signs such as that $(-)^{n-j}$ and $(-)^{q-l}$
 in \Eq{rhondown_para} and $(-)^q$ in \Eq{rhonup_para}
 result from the required time--ordering rearrangements,
 together with the Grassmann anticommutation relation;
 see \App{thapp_theo2} for details.

  The $n$ indexes in $\rho_{\ind}$,
 as specified in \Eq{rhon_para},
 should all be distinct, due to the Grassmann anticommutation
 relation. As results,
 the hierarchy in \Eq{final_para} would be finite,
 provided that the exponential series
 of reservoir correlation functions
 [\Eq{corr_para}] is effectively finite, such as
 the high--temperature cases.
 At zero temperature, the number of Matsubara terms required
 goes to infinity, and the HEOM via the present
 parametrization scheme fails.
 Nevertheless, the MFD--HEOM formalism in \Eq{final_theo}
 is remains valid, despite the cost due to the multi--dimensional
 frequency integration.
  Like its frequency--dispersed counterpart,
 the $n^{\rm th}$--tier ADO, $\rho_{\ind}$,
 is of the $(2n)^{\rm th}$--order system--reservoir
 coupling for its leading term.
 The hierarchy truncation can therefore be
 done in a similar manner, such as
 setting all $\rho^{(n>N_{\rm trun})}\approx 0$,
 followed by a convergency test.
 Apparently, the transport current can be readily
 expressed in terms of the first--tier ADO [cf.\ \Eq{currI}].

 \section{Concluding remarks}
 \label{thsum}

   We have established  the HEOM formalism, via both
 the MFD and the parametrization schemes
 (\Sec{ththeo} and \Sec{ththeo2}), for the dynamics
 of a general electron/spin system
 in contact with electrodes.
 It provides a unified tool to the study
 of a variety of quantum transport behaviors.
 These include the effects of Coulomb interaction,
 time--dependent electric potentials (external fields) applied
 on electrodes (system),
 multiple--terminals with different temperatures,
 and non-Markovian reservoir couplings
 on transport current.

   It is easy to show that the commonly
 used second--order QDT can be recovered
 with the first-tier truncation,
 while various fourth--order theories, such
 as the Liouville equation
 in Ref.\ \onlinecite{Ped05195330}
 are of the second--tier approximation here.
 In particular, we have demonstrated
 explicitly
 that the real-time diagrammatic
 formalism\cite{Sch9418436,Sch94423,Kon9531,Kon9616820,Kon961715}
 that has been used in the study of Kondo physics
 in quantum transport systems
 amounts to the second--tier truncation
 of the present HEOM formalism; see \Sec{thtier2}.

  The present theory is in principle
 exact, as the only approximation involved,
 the initial factorization ansatz in \Eq{rhoT0},
 can be removed by setting
 the initial time $t_0$ to infinite past.
 Therefore, at any given finite time before
 the application of time-dependent external fields,
 say $t=0$, the reduced system together
 with its grand canonical bath environment
 are in a steady state.
 This is determined as the
 steady-state solutions to the MFD-HEOM formalism,
 \Eq{final_theo} for general cases or
 \Eq{final_green} for single-particle systems,
 at either equilibrium if $\mu_{\alpha}=\mu^{\rm eq}_{\alpha}$,
 or nonequilibrium if $\mu_{\alpha}\neq\mu^{\rm eq}_{\alpha}$
  but time independent.
 Not only to the reduced system density operator,
 the steady--state solutions are
 also to those bath--induced auxiliary ones.
  They carry all relevant information on
 the correlations between system and reservoirs,
 as dictated by the HEOM theory.
 The resulting stationary solutions
 are used as the initial conditions
 at $t=0$. The subsequent reduced dynamics
 and transient transport properties
 are then evaluated via the present formalism
 again, upon switch-on of time-dependent $\Delta_\alpha(t)$,
 in additional to the constant
 $\mu_{\alpha}$ at earlier time.
 Consequently, the present hierarchical QDT  formalism
 is exact, without any approximation.

  The present theory recovers exactly the
 Landauer--B\"{u}ttiker's transport current
 expression, as it should.
 The resulting RSPDM-HEOM [\Eq{final_green}],
 which is exact for a single--particle system,
 is particular appealing due to its numerical feasibility
 for large systems. It may lead to a practical scheme of the
 time-dependent density functional theory (TDDFT)
 for open many-particle systems. In principle,
 this can be done by combining the present RSPDM-HEOM with the
 conventional DFT.\cite{Hoh64B864,Koh65A1133,Run84997}
 It is anticipated that the single-particle
 $h$-matrix in \Eq{final_green} be mapped
 to the Kohn-Sham
 counterpart,\cite{Yam03153105,Li07075114,Zhe07195127,Ven07226403}
 \be\label{hKS}
   h_{\mu\nu}(t)=h^0_{\mu\nu} + v^{\rm xc}_{\mu\nu}(t)
    + \sum_{\mu'\nu'}\varrho_{\mu'\nu'}(t)V_{\mu\nu\mu'\nu'},
 \ee
 where $h^0$ is the single--electron contribution and
 $V_{\mu\nu\mu'\nu'}$ the two--electron Coulomb integral.
 The key issue is how to identify
 the exchange--correlation potential $v^{\rm xc}_{\mu\nu}(t)$ in
 the TDDFT for open many--particle systems.
 The exact HEOM formalism, which is numerically
 feasible for model Coulomb--interaction systems,
 may shed some light on the construction
 of exchange--correlation functionals.

\begin{acknowledgments}
 Support from the RGC (604007)
 of Hong Kong is acknowledged.
\end{acknowledgments}

\appendix
\section{Path integral formalism: derivation}
\label{thapp_path}
   The formal solution to the total density
 operator in the $h_{\B}$-interaction is
 \be\label{app_path_rhoT}
  \rho_{\rm T}(t)=U_{\rm T}(t,t_0;\{\hat f^\sigma_{\alpha\mu}(t)\})
   \rho_{\rm T}(t_0)U^\dg_{\rm T}(t,t_0;\{\hat f^\sigma_{\alpha\mu}(t)\}),
 \ee
 with $U_{\rm T}(t,t_0;\{\hat f^\sigma_{\alpha\mu}(t)\})$
  being the stochastic Hilbert-space propagator, satisfying
 $\partial_t U_{\rm T} = -i[H+H'(t)]U_{\rm T}$.
 Let $\{|\psi\ra\}$ be a second-quantization
 basis set in the system subspace.
 The PI expression of $U_{\rm T}$ reads
 \begin{align} \label{pathUT}
   &\quad U_{\rm T}(\psi,t;\psi_0,t_0;\{\hat f^\sigma_{\alpha\mu}(t)\})
 \nl&=
   \int_{\psi_0}^{\psi}\!\!{\cal D}{\psi}\, e^{iS[\psi]}
  \exp_{+}\bigg\{
    -i\sum_{\alpha\mu} \!\int_{t_0}^{t}\!d\tau
% \nla\qquad \times
   \Bigl(a_{\mu}[\psi(\tau)]\hat f^\dg_{\alpha\mu}(\tau)
 \nla\qquad\qquad\qquad\qquad\qquad
  +\hat f_{\alpha\mu}(\tau)a^\dg_{\mu}[\psi(\tau)]\Bigr)\bigg\} .
 \end{align}
%%%
 The action functional $S[\psi]$ is related to the
 isolated system Hamiltonian only;
 $\{a^{\sigma}_{\mu}[\psi(\tau)]\}$ are
 the Grassmann variables,\cite{Ryd96,Kle06}
 as they denote the creation/annihilation operator of system
 in the Fermion field PI representation.
  On the other hand, the stochastic bath variables
 $\{\hat f^{\sigma}_{\alpha\mu}(t)\}$
 remain as the original operators, for which
 the time-ordered exponential function is needed.

   Consider now the reduced system density matrix
 $\rho(t)\equiv {\rm tr}_{\B}[\rho_{\rm T}(t)]$.
 Using \Eq{app_path_rhoT}, together
 with the initial factorization ansatz of \Eq{rhoT0},
 it is obtained that  (setting $\beta = \beta_{\alpha}$
 for simplicity)
\bsube\label{rhot_U}
 \begin{align} \label{rhot_Ua}
  \rho(t) &= {\rm tr}_{\B}
   \big[U_{\rm T}(t,t_0;\{\hat f^\sigma_{\alpha\mu}(t)\})
     \rho_{\rm T}(t_0)
    U^{\dg}_{\rm T}(t,t_0;\{\hat f^\sigma_{\alpha\mu}(t)\})\big]
 \nl&=
   {\rm tr}_{\B}\big[e^{\beta(h_{\B}-{\bm\mu}\hat{\bm N})}
    U_{\rm T}(t,t_0;\{\hat f^\sigma_{\alpha\mu}(t)\})
    e^{-\beta(h_{\B}-{\bm\mu}\hat{\bm N})}
 \nla \qquad\quad \times
   \rho(t_0)U^{\dg}_{\rm T}(t,t_0;\{\hat f^\sigma_{\alpha\mu}(t)\})
   \rho^{0}_{\B}\big ]
 \nl&=
  \big\la U_{\rm T}(t,t_0;\{\ti f^\sigma_{\alpha\mu}(t-i\beta)\})
\nla \qquad\quad \times
  \rho(t_0)
   U^\dg_{\rm T}(t,t_0;\{\hat f^\sigma_{\alpha\mu}(t)\}) \big\ra_{\B},
 \end{align}
 with ${\bm\mu}\hat{\bm N} \equiv \sum_{\alpha}\mu_{\alpha}\hat N_{\alpha}$
 and
 \begin{align}\label{app_tildeF}
   \ti f^\sigma_{\alpha\mu}(t-i\beta)
  &\equiv
     e^{\beta(h_{\B}-{\bm\mu}\hat{\bm N})}
     \hat f^\sigma_{\alpha\mu}(t)
     e^{-\beta(h_{\B}-{\bm\mu}\hat{\bm N})}
 \nl&=
     e^{-\sigma\beta\mu_\alpha}\hat f^\sigma_{\alpha\mu}(t-i\beta).
 \end{align}
 \esube
In writing the last identity, the relations of
$[\hat N_{\alpha},h_{\B}]=0$ and
 $e^{-\beta{\bm\mu}\hat{\bm N}}
  f^\sigma_{\alpha\mu}e^{\beta{\bm\mu}\hat{\bm N}}
 = e^{-\sigma\beta\mu_\alpha}f^\sigma_{\alpha\mu}$
 are used,\cite{Jin07134113} together with \Eq{falpt}.

  The influence functional used in \Eq{U0_def}
 can then be evaluated by using \Eqs{pathUT} and
 (\ref{rhot_U}), together with the Gaussian
 statistics for the stochastic bath operators $\{\hat f_{\alpha\mu}(t)\}$.
 The details are as follows.
 \begin{align} \label{app_FV}
  {\cal F}&=
    \Bigg\la
      \exp_{+}\biggl\{-i\sum_{\alpha\mu}\!\int_{t_0}^{t}\!d\tau
    \Big(e^{-\beta\mu_\alpha}
        \hat f^{\dg}_{\alpha\mu}(\tau-i\beta) a_{\mu}[\psi(\tau)]
  \nla\qquad\qquad\qquad\qquad
    +e^{\beta\mu_\alpha}
      a^\dg_{\mu}[\psi(\tau)]\hat f_{\alpha\mu}(\tau-i\beta)
     \Big)\biggr\}
  \nla \times
     \exp_{-}\biggl\{i\sum_{\alpha\mu}\!\int_{t_0}^{t}\!d\tau
     \Big(
       \hat f^{\dg}_{\alpha\mu}(\tau)a_{\mu}[\psi'(\tau)]
  \nla\qquad\qquad\qquad\qquad
     + a^\dg_{\mu}[\psi'(\tau)]\hat f_{\alpha\mu}(\tau)
     \Big)\biggr\}
 \Bigg\ra_{\B} .
 \end{align}
 For $\{\hat f^{\sigma}_{\alpha\mu}(t)\}$
 satisfying Gaussian statistics, the
 bath ensemble average in \Eq{app_FV} can be evaluated exactly by
 using the second-order cumulant expansion method, as the higher
 order cumulants are all zero.
 This property, together with \Eq{ctsym}, leads to
 the influence exponent in ${\cal F}\equiv \exp(-\Phi)$
 the following expression,
 \begin{align}\label{app_Phi}
  &\quad \Phi[\psi,\psi']
 \nl&= \sum_{\alpha\mu\nu} \!
    \int^t_{t_0}\!\!d\tau_2\!\!\int^{\tau_2}_{t_0}\!\!d\tau_1
  \Big\{a_\mu[\psi(\tau_2)]a^{\dg}_\nu[\psi(\tau_1)]
    C^{+}_{\alpha\mu\nu}(\tau_2-\tau_1)
 \nl&\qquad\qquad
  +a^{\dg}_\mu[\psi(\tau_2)]a_\nu[\psi(\tau_1)]
   C^{-}_{\alpha\mu\nu}(\tau_2-\tau_1)
 \nl&\qquad\qquad
   +a_\nu[\psi'(\tau_1)]a^{\dg}_\mu[\psi'(\tau_2)]
    C^{+\ast}_{\alpha\mu\nu}(\tau_2-\tau_1)
 \nl&\qquad\qquad
   +a^{\dg}_\nu[\psi'(\tau_1)]a_\mu[\psi'(\tau_2)]
    C^{-\ast}_{\alpha\mu\nu}(\tau_2-\tau_1)
  \Big\}
\nla
  -\sum_{\alpha\mu\nu}\!
    \int^t_{t_0}\!\!d\tau_2\!\!\int^{t}_{t_0}\!\!d\tau_1
    \Big\{a_\mu[\psi(\tau_2)]a^{\dg}_\nu[\psi'(\tau_1)]
     C^{-\ast}_{\alpha\mu\nu}(\tau_2-\tau_1)
 \nl&\qquad\qquad
    +a^{\dg}_\mu[\psi(\tau_2)]a_\nu[\psi'(\tau_1)]
     C^{+\ast}_{\alpha\mu\nu}(\tau_2-\tau_1)\Big\} .
 \end{align}
%%%%%%%%%%%%%
 Here, we have used the symmetry relation
 [the first identity of \Eq{ctsym}] in
 the third and fourth terms, and the
 detailed-balance relation [the second identity of \Eq{ctsym}]
 in the last two terms of the above expression.
 Some elementary algebra will then lead to \Eq{app_Phi},
 recast in terms of ${\cal R}\equiv \partial_t \Phi$, the expression,
 \begin{align}\label{app_calR}
    {\cal R}[t;\{\bm\psi\}]
 &= \sum_{\alpha\mu\sigma}
    \Big(a^{\bar\sigma}_\mu[\psi(t)]
      \big\{B^\sigma_{\alpha\mu}(t;\{\psi\})
               -B'^{\sigma}_{\alpha\mu}(t;\{\psi'\})
      \big\}
 \nl&\qquad
    -\big\{B^\sigma_{\alpha\mu}(t;\{\psi\})
      -B^{\prime\sigma}_{\alpha\mu}(t;\{\psi'\})\big\}
    a^{\bar\sigma}_\mu[\psi'(t)]\Big)
 \nl&=
    i\sum_{\alpha\mu\sigma} \Big\{
    a^{\bar\sigma}_\mu[\psi(t)]
    {\cal B}^{\sigma}_{\alpha\mu}\big(t;\{\bm\psi\}\big)
 \nl&\qquad\qquad
   -{\cal B}^{\sigma}_{\alpha\mu}\big(t;\{\bm\psi\}\big)
   a^{\bar\sigma}_\mu[\psi'(t)]
   \Big\}.
 \end{align}
 Here,  ${\cal B}^{\sigma}_{\mu} $ and $B^{\sigma}_{\mu} $
 ($B'^{ \sigma}_{\mu} $) are the same
 Grassmann variables defined in \Eq{calBs} and \Eq{calBs0}, respectively.
 In particular, ${\cal B}^{\sigma}_{\alpha\mu}
   a^{\bar\sigma}_\mu[\psi'(t)]
 = -a^{\bar\sigma}_\mu[\psi'(t)]
 {\cal B}^{\sigma}_{\alpha\mu}$,
 since $a^{\bar\sigma}_\mu[\psi'(t)]$ is also a
 Grassmann variable. With
 ${\cal A}_{\mu}^{\sigma}$ defined in \Eq{calAs},
 the above equation is identical to \Eq{calF_R}.

\section{Derivation of EOM (\ref{final_green})
    for single-particle Hamiltonian systems}
\label{thapp_green}

  Applying the primary-tier MFD-HEOM, i.e.,
 \Eq{final_theo} with $n=0$, for the RSPDM of \Eq{DM0} leads to
 \begin{align} \label{app_dotrho1}
   i\dot \varrho_{\mu\nu}(t)
   &=
    {\rm tr}_{\rm s} \big[(a^{\dg}_{\nu}a_{\mu}{\cal L})\rho(t)\big]
 \nla
   +\sum_{\alpha m\sigma} \int d\omega\,  {\rm tr}_{\rm s}
    \big[(a^{\dg}_{\nu}a_{\mu}{\cal A}^{\bar\sigma}_{m})
         \phi^{\sigma}_{\alpha m}(\omega,t) \big]
 \nl&=
    {\rm tr}_{\rm s} \big\{[a^{\dg}_{\nu}a_{\mu},H]\rho(t)\big\}
 \nla
   +\sum_{\alpha m\sigma} \int d\omega\,  {\rm tr}_{\rm s}
    \big\{[a^{\dg}_{\nu}a_{\mu},a^{\bar\sigma}_m]
         \phi^{\sigma}_{\alpha m}(\omega,t) \big\}.
 \end{align}
 The second identity arises from the trace cyclic invariance.
 For the single-particle system of \Eq{H1e},
 \be \label{aaH}
    [a^{\dg}_{\nu}a_{\mu},H]
   =\sum_m (h_{\mu m}a^{\dg}_{\nu}a_m - h_{m\nu}a^{\dg}_m a_{\nu}).
 \ee
 It leads to
 \be\label{aaL}
   {\rm tr}_{\rm s} \big[(a^{\dg}_{\nu}a_{\mu}{\cal L})\rho(t)\big]
    = [h,\varrho]_{\mu\nu}.
 \ee
 Next,
 $[a^{\dg}_{\nu}a_{\mu},a_m]=-a_{\mu} \delta_{\nu m}$
 and $[a^{\dg}_{\nu}a_{\mu},a^{\dg}_m]=a^{\dg}_{\nu}\delta_{\mu m}$
 lead to
 \begin{align}
    & \sum_{\sigma m} {\rm tr}_{\rm s}
      \big\{[a^{\dg}_{\nu}a_{\mu},a^{\bar\sigma}_m]
         \phi^{\sigma}_{\alpha m}(\omega,t) \big\}
 \nl=&
      -{\rm tr}_{\rm s}[a_{\mu}\phi^{+}_{\alpha\nu}(\omega,t)]
       +{\rm tr}_{\rm s}[a^{\dg}_{\nu}\phi^{-}_{\alpha\mu}(\omega,t)]
 \nl\equiv&
      -\varphi^{+}_{\alpha\mu\nu}(\omega,t)
      +\varphi^{-}_{\alpha\nu\mu}(\omega,t),
 \end{align}
 with
 \bsube \label{varphi_pm}
 \begin{align}
   \varphi^{+}_{\alpha\mu\nu}(\omega,t)
 &\equiv
     {\rm tr}_{\rm s}[a_{\mu}
    \phi^{+}_{\alpha\nu}(\omega,t)] = \varphi_{\alpha\mu\nu}(\omega,t),
 \label{varphi_plus} \\
   \varphi^{-}_{\alpha\nu\mu}(\omega,t)
 &\equiv
     {\rm tr}_{\rm s}[a^{\dg}_{\nu}
    \phi^{-}_{\alpha\mu}(\omega,t)]
  = \varphi^{\ast}_{\alpha\nu\mu}(\omega,t).
 \label{varphi_minus}
 \end{align}
 \esube
 The second identity in \Eq{varphi_plus} is the same as \Eq{DM1}.
 In writing \Eq{varphi_minus}, the property of
 $\phi^{\sigma}_{\alpha\mu}=[\phi^{\bar\sigma}_{\alpha\mu}]^{\dg}$
 that leads to $\varphi^{\bar\sigma}_{\alpha}(\omega,t)
  = [\varphi^{\sigma}_{\alpha}(\omega,t)]^{\dg}$
 is used.
  We have thus arrived at \Eq{final_green0}, which
 is just the matrix form of \Eq{app_dotrho1}.

   We are now in the position to show \Eq{final_green1}.
 The MFD-HEOM [\Eq{final_theo}]
 for the first-tier auxiliary RSPDM in \Eq{DM1} reads
 \begin{align}\label{app_dotvarphi1}
   i\dot\varphi_{\alpha\mu\nu}
 &=
   {\rm tr}_{\rm s}[(a_{\mu}{\cal L})\phi^{+}_{\alpha\nu}]
  - (\omega+\Delta_{\alpha})\varphi_{\alpha\mu\nu}
 \nla
   + {\rm tr}_{\rm s}
     \left\{\big[a_{\mu}{\cal C}^{+}_{\alpha\nu}(\omega)\big]\rho\right\}
 \nla
   +\int d\omega'
    \sum_{\alpha'} \varphi_{\alpha'\mu,\alpha\nu}(\omega',\omega,t),
 \end{align}
 with
 \be \label{app_green_varphi2_def}
   \varphi_{\alpha'\mu,\alpha\nu}
  \equiv \sum_{\sigma,m} {\rm tr}_{\rm s}
    \big[(a_{\mu}{\cal A}^{\bar\sigma}_{m})
    \phi^{+,\sigma}_{\alpha\nu,\alpha'm}(\omega,\omega',t)\big].
 \ee
   The first term in the rhs of \Eq{app_dotvarphi1}
 can be evaluated by using the identity,
  $a_{\mu}{\cal L}=[a_{\mu},H]= \sum_m h_{\mu m}a_m$,
 for the present single-particle system, together
 with \Eq{DM1}. It results in
 \be\label{aLphi1}
  {\rm tr}_{\rm s}[(a_{\mu}{\cal L})\phi^{+}_{\alpha\nu}]
  = \sum_m \big(
    h_{\mu m} {\rm tr}_{\rm s}[a_m\phi^{+}_{\alpha\nu}]\big)
  = \big(h\varphi_{\alpha}\big)_{\mu\nu}.
 \ee
  The third term in the rhs of \Eq{app_dotvarphi1}
 can be evaluated as [cf.\ \Eqs{CAphiB} and (\ref{DM1})]
 \begin{align*}
  &\quad
    {\rm tr}_{\rm s}
  \left\{\big[a_{\mu}{\cal C}^{+}_{\alpha\nu}(\omega)\big]\rho\right\}
 \nl&=
  \sum_m {\rm tr}_{\rm s}
  \left[\Gamma^{+}_{\alpha\nu m}(\omega)a_{\mu}a^{\dg}_m \rho
      - \Gamma^{-}_{\alpha m\nu}(\omega)a^{\dg}_m a_{\mu}\rho
  \right]
 \nl&=
   \sum_m \Gamma^{+}_{\alpha\nu m}(\omega)\bar\varrho_{\mu m}
     - \Gamma^{-}_{\alpha m\nu}(\omega)\varrho_{\mu m}.
 \end{align*}
 Here, $\bar\varrho_{\mu m}\equiv {\rm tr}_{\rm s}(a_{\mu}a^{\dg}_m \rho)
  = \delta_{\mu m}-\varrho_{\mu m}$; i.e.,
 the elements of reduced single--hole density matrix of
 $\bar \varrho\equiv{\bf 1}-\varrho$.
 Together with \Eqs{Jwsym} and  (\ref{FDTw1}), the
 above equation can be recast as
 \be \label{app_green_aCrho}
   {\rm tr}_{\rm s}
  \left\{\big[a_{\mu}{\cal C}^{+}_{\alpha\nu}(\omega)\big]\rho\right\}
 = \big[f_{\alpha}(\omega)J_{\alpha}(\omega)
  - \varrho J_{\alpha}(\omega)\big]_{\mu\nu}.
 \ee

  Note that due to the Grassmann parity of \Eq{CAphiA},
  ${\cal A}^{\bar\sigma}_{m}$ in \Eq{app_dotrho1}
 behaves as a commutator, its action
 in \Eq{app_green_varphi2_def} is an anticommutator.
 The fact that $\varphi_{\alpha'\mu,\alpha\nu}$ defined
 in \Eq{app_green_varphi2_def} is identical
 to the matrix
 element $[\varphi_{\alpha'\alpha}(\omega',\omega,t)]_{\mu\nu}$
 of \Eq{DM2} can then be readily concluded.
  We have thus completed \Eq{final_green1}, with the
 $S_{\alpha}(\omega)$ defined in \Eq{S_def}.

  Finally, \Eq{final_green2} can be readily obtained
 via the trace of the second-tier MFD-HEOM [\Eq{final_theo}],
 together with \Eqs{Jwsym}, (\ref{varphi_pm}),
 and the elementary algebra just described.

   Note that the coupled EOM, \Eq{final_green}, can formally be recast
 in the standard form as
 \bsube\label{app_final_green}
 \be \label{app_final_green0}
  i\frac{\partial}{\partial t}
  \begin{bmatrix}
   {\sf X} \\ {\sf Y}
  \end{bmatrix}
 =
  [{\Lambda}+\delta{\Lambda}(t)]
  \begin{bmatrix}
   {\sf X} \\ {\sf Y}
  \end{bmatrix}
 +
  \begin{bmatrix}
   {\sf 0} \\ {\sf S}
  \end{bmatrix},
 \ee
 with
 \be \label{app_green_XY}
   {\sf X} \equiv
    \begin{bmatrix}
      \bm\varrho(t)  \\  \bm\varphi_{\alpha'\alpha}(\omega',\omega,t)
    \end{bmatrix};
 \ \ \
  {\sf Y}\equiv
   \begin{bmatrix}
    \bm\varphi_{\alpha}(\omega,t) \\ \bm\varphi^\dg_{\alpha'}(\omega',t)
   \end{bmatrix},
 \ee
 and
 \be \label{app_greenV}
  {\sf S} \equiv \begin{bmatrix}
    {\bm S}_{\alpha}(\omega) \\ -{\bm S}_{\alpha'}(\omega')
   \end{bmatrix}.
 \ee
 \esube
 The $4\times 4$ matrix $\delta{\Lambda}(t)$ in \Eq{app_final_green0}
 arises from the time-dependent bias potential.
 It is diagonal, with the elements of
 \be\label{app_green_Det}
  \delta{\Lambda}(t)= \{0,-\Delta_{\alpha}(t)+\Delta_{\alpha'}(t),
    -\Delta_{\alpha}(t), \Delta_{\alpha'}(t)\}.
 \ee
 The time-independent counterpart is given in terms of
 block-matrix form as
 \bsube \label{app_greenABC}
 \be \label{app_greenABCa}
  {\Lambda} = \begin{bmatrix}
   \Lambda_{\sf xx} & \Lambda_{\sf xy}
  \\ \Lambda_{\sf yx} & \Lambda_{\sf yy}
  \end{bmatrix} ,
 \ee
 with
 \begin{align}
   \Lambda_{\sf xx} &=\begin{bmatrix}
      \leftrightact{h} & 0  \\  0 & \omega'-\omega
   \end{bmatrix},
 \ \quad
  \Lambda_{\sf xy} =\begin{bmatrix}
    -{\bm 1}_\alpha & {\bm 1}_{\alpha'}
     \\ \leftact{J}_{\alpha'} &-\rightact{J}_{\alpha}
   \end{bmatrix},
 \label{app_greenABCb} \\
  \Lambda_{\sf yx} &=\begin{bmatrix}
    -\rightact{J}_{\alpha} & {\bm 1}_{\alpha'}
    \\
    \leftact{J}_{\alpha'} & -{\bm 1}_{\alpha}
  \end{bmatrix},
 \ \
 \Lambda_{\sf yy} =\begin{bmatrix}
    \leftact h -\omega & 0 \\ 0 &  \omega'-\rightact h
  \end{bmatrix},
 \label{app_greenABCc}
 \end{align}
 \esube
%%%
 where
 \[
   {\bf 1}_{\alpha} \equiv \sum_{\alpha}\int d\omega,
 \ \ \
  {\bf 1}_{\alpha'} \equiv \sum_{\alpha'}\int d\omega'.
 \]
 Introduced in \Eq{app_greenABC}
 is also the tetradic notation for
 the left- or right-multiplication action of a matrix $O$
 on the matrix $A$ of interest as
 \[
   \leftact{O}A \equiv OA, \ \ \  \rightact{O}A \equiv AO^{\dg},
 \ \ {\rm and\ \ } \leftrightact{O} \equiv \leftact{O} - \rightact{O}.
 \]
 In other words, $\leftact{O}$ and $\rightact{O}$ are tensors, with the
 elements of $\leftact{O}_{mn,m'n'}=O_{mm'}\delta_{n'n}$
 and $\rightact{O}_{mn,m'n'}=\delta_{mm'}O^{\ast}_{nn'}$.
 Implied in \Eq{app_greenABC} is also
 the frequency variable associating with the subindex;
 i.e., $\rightact{J}_{\alpha} \equiv \rightact J_{\alpha}(\omega)$
 and $\leftact{J}_{\alpha'} \equiv \leftact J_{\alpha'}(\omega')$.

 \section{Derivation of HEOM (\ref{final_para})}
 \label{thapp_theo2}

    Denote ${\bf k}\equiv (\alpha\mu\nu\sigma k)$
 and ${\bf m}\equiv (\alpha\mu\sigma m)$ for short,
 and introduce
 \bsube \label{Bpara}
 \begin{align}
    B_{\bf k}(t;\{\psi\})
 &\equiv \int^t_{t_0}\!\!d\tau
    e^{-\theta_{\bf k}(t,\tau)}
    a^{\sigma}_{\nu}[\psi(\tau)],
 \label{Bs_para1}\\
    \check B_{\bf m}(t;\{\psi\})
  &\equiv \sum_{\nu}\!
   \check\eta^{\sigma}_{\alpha\mu\nu m}
    \!\! \int^t_{t_0}\!\!d\tau
     e^{-\check\theta^{\sigma}_{\alpha m}(t,\tau)}
     a^{\sigma}_{\nu}[\psi(\tau)],
 \label{Bs_para2}
 \end{align}
 \esube
 with $\theta_{\bf k}(t,t)
  =\check\theta^{\sigma}_{\alpha m}(t,t)=0$, and
 \bsube\label{phase_para}
 \begin{align}
    \partial_t\theta_{\bf k}(t,\tau)
 &= \gamma^{\sigma}_{\alpha\mu\nu k}-\sigma i\Delta_\alpha(t),
 \\
    \partial_t\check\theta^{\sigma}_{\alpha m}(t,\tau)
 &= \check\gamma^{\sigma}_{\alpha m}-\sigma i\Delta_\alpha(t).
 \end{align}
 \esube
 The two functionals in \Eq{Bpara}
 are the counterparts of that in \Eq{calBs0a}, arising
 from the two distinct components of
 reservoir correlation functions as \Eq{corr_para}.
 The nonstationary exponential factor
 in \Eq{corr2}, due to time--dependent
 chemical potentials applied to electrodes,
 are also accounted for via
 the nonstationary phases as \Eq{phase_para}.

  The dissipation functional [\Eq{calR}]
 is now decomposed according to the parametrization
 of \Eq{corr_para} as
 \be\label{app_calRpara}
  \mathcal{R}=i\sum_{\bf k}
   \mathcal{A}^{\bar\sigma}_{\mu}\mathcal{B}_{\bf k}
   + i\sum_{\bf m} \mathcal{A}^{\bar\sigma}_{\mu}
       \check {\cal B}_{\bf m},
 \ee
 which is the same as \Eq{calRpara},
 with [cf.\ \Eq{calBs}],
 \bsube \label{calBs_para}
 \begin{align} \label{calBk_para}
  {\cal B}_{\bf k}
 &\equiv
   -i\left[\eta^{\sigma}_{\alpha\mu\nu k}
   B_{\bf k}(t;\{\psi\})
      -\eta^{\bar\sigma\,\ast}_{\alpha\mu\nu k}
   B_{\bf k}(t;\{\psi' \})\right],
 \\ \label{calBm_para}
    \check {\cal B}_{\bf m}
 &\equiv -i\left[\check B_{\bf m}(t;\{\psi\})
         +\check B_{\bf m}(t;\{\psi'\})
      \right].
 \end{align}
 \esube
 In writing \Eq{calBm_para}, the last
 relation in \Eq{etaMat} is applied.
  The time derivatives on \Eq{calBs_para} can
 be obtained as [cf.\ \Eq{phase_para}]
 \bsube \label{dotcalB_para}
\begin{align}
   \partial_t {\cal B}_{\bf k}
 &=
   -[\gamma^{\sigma}_{\alpha\mu\nu k}-\sigma i\Delta_\alpha(t)]
      {\cal B}_{\bf k}
   -i{\cal C}_{\bf k},
 \\
   \partial_t \check {\cal B}_{\bf m}
 &=
   -[\check\gamma^{\sigma}_{\alpha m}-\sigma i\Delta_\alpha(t)]
     \check {\cal B}_{\bf m} -i\check{\cal C}_{\bf m},
\end{align}
\esube
 with
\bsube\label{app_calCPI_para}
\begin{align}
   {\cal C}_{\bf k}
 &= \eta^{\sigma}_{\alpha\mu\nu k}
   a^{\sigma}_{\nu}[\psi(t)]
      -\eta^{\bar\sigma\,\ast}_{\alpha\mu\nu k}
   a^{\sigma}_{\nu}[\psi'(t)] ,
\\
  \check{\cal C}_{\bf m}
 &= \sum_{\nu}\check\eta^{\sigma}_{\alpha\mu\nu m}
      \Big(a^{\sigma}_{\nu}[\psi(t)]+a^{\sigma}_{\nu}[\psi'(t)]
      \Big),
\end{align}
\esube
 in the PI representation.

   The $n^{\rm th}$--tier AIFs can in general be defined as
 \be\label{AIF_para}
  {\cal F}_{\ind} \equiv
  {\cal F}^{(n)}_{{\bf k}_1\cdots{\bf k}_p\!,{\bf m}_1\cdots{\bf m}_q}
  \equiv
    \check{\cal B}_{{\bf m}_q}\cdots\check{\cal B}_{{\bf m}_1}
    {\cal B}_{{\bf k}_p}\cdots{\cal B}_{{\bf k}_1}
     {\cal F},
 \ee
 with $n \equiv p + q$.
  The associated $n^{\rm th}$--tier ADO $\rho_{\ind}$
 is defined via (noting that $\rho \equiv \rho_{\sf 0}$)
 \bsube \be \label{rhosfn_def_para}
  \rho_{\ind}(t)\equiv
 {\cal U}_{\ind}(t,t_0)\rho(t_0),
 \ee
 with the propagator,
 \be \label{calGPIn_def}
   {\cal U}_{\ind}(t,\bm\psi;t_0,\bm\psi_0)
 \equiv \int_{{\bm\psi}_0[t_0]}^{{\bm\psi}[t]}
  {\cal D}{\bm\psi}
   \, e^{iS[\psi]} {\cal F}_{\ind} e^{-iS[\psi']} ,
 \ee
 \esube
 in the PI representation.

    The HEOM formalism [\Eqs{final_para}--(\ref{rhoup_para})]
 for $\rho_{\ind}$ can be followed immediately
 via the time derivative on ${\cal F}_{\ind}$.
 The latter is carried out by using \Eq{dotcalB_para}, together with
 $\partial_t {\cal F}=-{\cal R}{\cal F}$ and \Eq{app_calRpara}.
 Some details are as follows.

  The $\gamma_{\ind}(t)$ in \Eqs{final_para} and (\ref{gamma_n})
 arises from the square--bracket terms in the rhs of \Eq{dotcalB_para}.
 It collects the time derivatives on the nonstationary phases
 of \Eq{phase_para}.

  The second terms in \Eq{dotcalB_para} leads to
 the tier-down dependence, as shown by \Eq{rhondown_para},
 with the involving $(n-1)^{\rm th}$--tier ADOs of
 \Eq{rhodown_para} associating with
 \bsube \label{calFdown}
 \begin{align}
   {\cal F}_{\ind^{-}_{j}}
 &\equiv
    \check{\cal B}_{{\bf m}_q}\cdots\check{\cal B}_{{\bf m}_1}
    {\cal B}_{{\bf k}_p}\cdots{\cal B}_{{\bf k}_{j+1}}
    {\cal B}_{{\bf k}_{j-1}}\cdots{\cal B}_{{\bf k}_1}
 {\cal F},
 \label{calFdown_a}\\
   {\cal F}_{\check\ind^{-}_{l}}
 &\equiv
    \check{\cal B}_{{\bf m}_q}\cdots\check{\cal B}_{{\bf m}_{l+1}}
    \check{\cal B}_{{\bf m}_{l-1}}\cdots\check{\cal B}_{{\bf m}_1}
    {\cal B}_{{\bf k}_p}\cdots{\cal B}_{{\bf k}_1}
  {\cal F}.
 \label{calFdown_b}
 \end{align}
 \esube
  The ${\cal C}$--functionals in \Eqs{dotcalB_para}
 and (\ref{app_calCPI_para}), which depend only
 on the fixed ending points of PI at the
 local time $t$, are now superoperators
 in \Eqs{rhondown_para} and (\ref{calC_para}) with
 Grassmann parity.
 The signs $(-)^{n-j}$ and $(-)^{q-l}$
 in \Eq{rhondown_para}
 are associated with the time--ordering
 arrangement. It brings
 the ${\cal C}$ and
 $\check{\cal C}$ functionals, originally
 at the ${\bf k}_j$ and ${\bf m}_l$ positions,
 respectively, to the left most of action,
 as indicated in \Eq{calFdown}.

  The identity of $\partial_t {\cal F}=-{\cal R}{\cal F}$
 with \Eq{app_calRpara} contributes to $\partial_t {\cal F}_{\ind}$
 [\Eq{AIF_para}] the terms of
 \bsube\label{calFup}
 \begin{align}
  {\cal A}^{\bar\sigma}_{\mu}{\cal B}_{\bf k}{\cal F}_\ind
 &= (-)^{q}{\cal A}^{\bar\sigma}_{\mu}
   \left(\check{\cal B}_{{\bf m}_q}\cdots\check{\cal B}_{{\bf m}_1}
   {\cal B}_{\bf k}{\cal B}_{{\bf k}_p}\cdots{\cal B}_{{\bf k}_1}
   {\cal F}\right)
 \nl&
  \equiv (-)^{q}{\cal A}^{\bar\sigma}_{\mu}{\cal F}_{\ind^{+}_{\bf k}} ,
 \label{calFup_a}\\
   {\cal A}^{\bar\sigma}_{\mu}\check{\cal B}_{\bf m}{\cal F}_\ind
 &={\cal A}^{\bar\sigma}_{\mu}
   \left( \check{\cal B}_{\bf m}
    \check{\cal B}_{{\bf m}_q}\cdots\check{\cal B}_{{\bf m}_1}
    {\cal B}_{{\bf k}_p}\cdots{\cal B}_{{\bf k}_1}
      {\cal F}\right)
\nl& \equiv
   {\cal A}^{\bar\sigma}_{\mu}
   {\cal F}_{\check{\ind}^{+}_{\bf m}}.
 \label{calFup_b}
 \end{align}
 \esube
 They correspond to the two terms in \Eq{rhonup_para},
 respectively. The involving $(n+1)^{\rm th}$--tier
 ADOs, $\rho_{\ind^{+}_{\bf k}}$ and
 $\rho_{\check{\ind}^{+}_{\bf m}}$ [\Eq{rhoup_para}],
 are associated with ${\cal F}_{\ind^{+}_{\bf k}}$
 and ${\cal F}_{\check{\ind}^{+}_{\bf m}}$, respectively,
 defined in the second identities above.
 We have thus completed the derivation
 of the HEOM formalism via parametrization presented
 in \Sec{ththeo2}.

%\bibliography{/disk3/yan/refs/bibrefs}

\end{document}